\documentclass[A4,12pt]{article} 

\usepackage{ae}
\usepackage{multirow} 
\usepackage{graphicx}
\usepackage{subfigure}
\usepackage[round,authoryear]{natbib}
\usepackage{amsmath} 
\usepackage{amssymb}
\usepackage{indentfirst}
\usepackage[hang,small]{caption}
\usepackage{hyperref}
\hypersetup{colorlinks=true,linkcolor=blue,citecolor=blue,breaklinks={true}}

\newcommand{\bx}{\mbox{\boldmath $x$}}

\font\bbfnt=msbm10

\def\bbR{\mbox{\bbfnt R}}

\newcommand{\mb}[1]{\mbox{\bfseries \itshape #1}}
\newcommand{\1}{\Bar{1}} 
\newcommand{\2}{\Bar{\Bar{1}}}

\begin{document}

~\vspace{2.0cm}
\begin{center}
\begin{LARGE}
\bf
Observability  and Synchronization \vspace{0.40cm} \\   of Neuron Models \vspace{1.5cm} 
\end{LARGE}

{\sc  Luis A. Aguirre},   {\sc Leonardo L. Portes}  
 \vspace{0.25cm}

Departamento de Engenharia Eletr\^onica, Programa de P\'os Gradua\c{c}\~ao em
Engenharia El\'etrica da Universidade Federal de Minas Gerais, Av. Ant\^onio Carlos, 6627,
31.270-901 Belo Horizonte, MG, Brazil. {\tt aguirre@ufmg.br,~ll.portes@gmail.com}
 \vspace{0.5cm}

{\sc Christophe Letellier}
 \vspace{0.25cm}

Universit\'e de Rouen, CNRS, CORIA,UMR 6614, Campus Univ. Madrillet, F-76800 St Etienne, France. 
{\tt christophe.letellier@atomosyd.net}

\end{center}

%=======================================================
\section*{Abstract}

Observability is the property that enables to distinguish two different locations in $n$-dimensional
state space from a reduced number of measured variables, usually just one. In high-dimensional systems
it is therefore important to make sure that the variable recorded to perform the analysis conveys
good observability of the system dynamics. In the case of networks composed of neuron models,
the observability of the network depends nontrivially on the observability of the node dynamics and
on the topology of the network. The aim of this paper is twofold. First, a study of observability is
conducted using four well-known neuron models by computing three different observability coefficients.
This not only clarifies observability properties of the models but also shows the limitations of
applicability of each type of coefficients in the context of such models. Second, a
multivariate singular spectrum analysis (M-SSA) is performed to detect phase 
synchronization in
networks composed by neuron models. This tool, to the best of the authors' knowledge has not
been used in the context of networks of neuron models. It is shown that it is possible to
detect phase synchronization i)~without having to measure
all the state variables, but only one from each node, and ii)~without having to
estimate the phase.

%=======================================================
\section{Introduction}

Since the early days of last century there has been sustained activity in developing mathematical
models for neuron dynamics. More recently such models have been combined in networks in order to investigate
collective behavior. In either approaches mathematical tools and concepts abound, as reviewed
by \cite{sie_sta/16}, who argue that there must be a continued effort in using such tools to reveal
so many aspects of the brain dynamics which remain not understood. The same point had been argued
by Brown in a very intertaining discussion \citep{bro/14}.

In this respect, an important concept is that of observability of the dynamics from a given
measured variable. Since it is not practical, especially in high-dimension systems, to record all
state variables, a relevant problem is to know which are the best variables to record to be able to
infer the state of the whole system. Observability, although not in its classical interpretation, 
provides an answer to that question. It has been acknowledged that to choose variables that
provide good observability of the dynamics enables estimating the state of a network of 
neuron models using Kalman-related methods \citep{sed_eal/12,sch/12}. In a recent study about 
controllability and observability of network motifs built with neuron models, it has been
found that ``it is necessary to take the node dynamics into consideration when selecting
the best driver (sensor) node to modulate (observe) the whole network activity'' \citep[Sec.\,III-A]{su_eal/17}.

In view of this, one of the aims of this paper is to conduct a study of 
observability properties of four neuron models following three points of view: 
using the model equations and numerical analysis \citep{let_eal/05pre}, using 
the model equations and symbolic manipulations \citep{let_agu/09pre} and using 
time series data \citep{agu_let/11}. An interesting point that has been 
revealed in this study is related to aspects that are specific to neuron 
models. For instance, in the case of the Hodgkin-Huxley model, three of the 
four state variables are not directly measurable. The study of observability
could help understand if there are any serious limitations related to this. 
In such a case the use of observability coefficients estimated from data is 
most convenient, because the practical relevance of measuring ionic currents
could be evaluated. Due to the functional relation of such currents with the 
state variables, the computation of observability coefficients from the 
equations is significantly more difficult. Other examples are the integrate and 
fire models, that produce discontinuities in the data. Such phenomenon may have
adverse effects on data-driven observability coefficients, and the 
equation-based computation of coefficients is also questionable because of the 
``hidden state variable'' related to the firing process. These aspects, 
that have come to light in the context of the investigated neuron models, are 
here described for the first time. 

Another important aspect that has gained considerable attention is that of synchronization of
networked neuron models. Because in real life neurons are not identical and coupling could be weak,
phase synchronization is somewhat more well suited than complete synchronization in this context.
A difficulty with most procedures used to detect phase synchronization is the need for defining a
phase, which is not always simple, if at all possible. Spectral coherence related measures of phase synchronization
have recently been considered and found to deviate considerably from expected results \citep{low_eal/16}.
An alternative procedure that does not require the estimation of the phase is the multivariate spectrum 
analysis for phase synchronization phenomena, originally proposed by Groth and Ghill \citep{gro_ghi/11}.
This method will be reviewed and applied to detect phase synchronization in 
networks of neuron models in this work for the first time, to the best of the 
authors' knowledge. Although observability and synchronizability are different 
problems and treated as such in this paper, there is a connection between them 
in the context of multivariate spectrum analysis, as will be pointed out.

This paper is organized as follows. For the sake of completion, there are two sections with background
material. Section~\ref{nm} briefly describes four of the neuron models considered in this study.
Section~\ref{tools} reviews the main tools used: three different measures of observability and the
multivariate singular spectrum analysis, used in detecting phase synchronization. The numerical
results concerning observability of the investigated neuron models are presented in Sec.\,\ref{nr}
and the results regarding synchronization are briefly described in Sec.\,\ref{snnm}.
Conclusions are provided in Sec.\,\ref{conc}.

%=======================================================
\section{Neuron Models}
\label{nm}

This section surveys four neuron models presented in chronological order. 
This choice is admitedly {\it ad hoc}\, but it is believed
that some of the more commonly used neuron models are included. 
In presenting the equations, the symbols used in the
original publications have been maintained whenever possible. A general comparison of the
models used in this paper and many other is provided in \citep{izh/04}.

%----------------------------------------------------------------------------
\subsection{Hodgkin-Huxley Model}

In 1952, Alan Hodgkin and Andrew Huxley published a series of four papers that 
concluded with a biophysically-based model of neuron dynamics, known as the 
Hodgkin-Huxley model \citep{hod_hux/52}:
\begin{eqnarray}
\label{HH52}
\left\{ 
\begin{array}{l}
\dot{V}  =  \frac{1}{C_{\rm m}} \left( I -I_{\rm K} -I_{\rm Na} - I_l \right) \\
\dot{n}  =  \alpha_n(1-n)-\beta_n n  \\
\dot{m}  =  \alpha_m(1-m)-\beta_m m \\
\dot{h}  =  \alpha_h(1-h)-\beta_h h ,
\end{array} \right.
\end{eqnarray}
where
\begin{eqnarray}
I_{\rm K} & = & \bar{g}_{\rm K}n^4(V-V_{\rm K}),~ I_{\rm Na} = \bar{g}_{\rm Na}m^3h(V-V_{\rm Na}),~I_l  =  \bar{g}_l(V-V_l) \nonumber \\
\alpha_n & = & \displaystyle \frac{0.01(V+10)}{e^{\frac{V+10}{10}}-1},~\beta_n =  \displaystyle 0.125 e^{\frac{V}{80}} \nonumber\\
\alpha_m & = & \displaystyle \frac{0.1(V+25)}{e^{\frac{V+25}{10}}-1},~\beta_m = \displaystyle 4 e^{\frac{V}{18}} \nonumber\\
\alpha_h & = &  \displaystyle 0.07 e^{\frac{V}{20}},~\beta_h=\displaystyle \left[ e^{\frac{V+30}{10}}+1 \right]^{-1} , \nonumber 
\end{eqnarray}

\noindent
with the following parameter values: membrane capacitance 
$C_{\rm m}=1\,\mu{\rm F/cm}^2$; constant membrane potentials 
$V_{\rm K}=12$\,mV, $V_{\rm Na}=-115$\,mV, $V_l=-10.6$\,mV; constants 
associated with membrane conductances $\bar{g}_{\rm K}=36{\rm mS/cm}^2$, 
$\bar{g}_{\rm Na}=120{\rm mS/cm}^2$ and the conductance 
$\bar{g}_l=0.3{\rm mS/cm}^2$. $I$ is the total current density through the 
membrane, and $I_{\rm K}$, $I_{\rm Na}$ and $\bar{g}_l$ correspond to the 
current density due to potassium ions, sodium ions and other ions, 
respectively. All current densities are given in $\mu{\rm A/cm}^2$. 
Variables $n$, $m$ and $h$ are dimensionless variables corresponding 
to the proportion of the potassium inside of the membrane, the proportion of
activating molecules within the membrane, and the proportion of inactivating 
molecules outside of it, respectively.
% rkHH52.m; dvHH52.m; simHH52RK.m

%----------------------------------------------------------------------------
\subsection{FitzHugh-Nagumo Model}

In 1961 Richard FitzHugh published a model obtained from the van der Pol's equation
\citep{fit/61}:
\begin{eqnarray}
\label{FN61}
\left\{ 
\begin{array}{l}
\dot{x} =  c(y+x-x^3/3+I)  \\
\dot{y} = -(x-a+by)/c  ,
\end{array} \right.
\end{eqnarray}

\noindent
where $(a,\,b,\,c)$ are constant parameters and $I$ is a stimulus, that corresponds to membrane
current in the Hodgkin-Huxley model. $x$ is usually identified with membrane potential, and
$y$ is the recovery variable.
In the following year, Jin-Ichi Nagumo and colleagues published an 
electronic implementation of model (\ref{FN61}) that used tunnel diodes \citep{nag_eal/62}, hence
 (\ref{FN61}) is often referred to as the FitzHugh-Nagumo model \citep{sch/12} and it is considered to be
a simplified version of the Hodgkin-Huxley model in the sense that it reproduces some of the main features
of the dynamics. 

Model (\ref{FN61}) with $(a,\,b,\,c)=(0.7,\,0.8,\,3)$ was investigated in \citep{fit/61} with $I=-0.4$ and in
\citep{voo_eal/04} with $I(t)$ being a slowly varying stimulus within the range $-1.5 \le I(t) \le -0.4$.
% rkFN61.m; dvFN61.m; simFN61RK.m

%----------------------------------------------------------------------------
\subsection{Hindmarsh-Rose Model}
\label{mod_HR}

In 1984 Hindmarsh and Rose provided a modification to a previous model published by them in 1982 thus
yielding the following equations for a three-equilibrium-point model with adaptation \citep{hin_ros/84}
\begin{eqnarray}
\label{HR84}
\left\{ 
\begin{array}{l}
\dot{x}  =  y -ax^3+b x^2+I -z  \\
\dot{y}  =  c - d x^2 -y  \\
\dot{z}  =  r[s(x-x_1)-z] ,  
\end{array} \right.
\end{eqnarray}

\noindent
with $(a,\,b,\,c,\,d)=(1,\,3,\,1,\,5)$, where $x_1$ is a  constant. 
The constant parameters 
$r$ and $s$ determine the dynamical response to a short pulse of depolarizing current. 
Here $x$ is the membrane potential and $y$ is the recovery variable (as in the FitzHugh-Nagumo model). In this model
Hindmarsh and Rose added a third equation, where $z$ is an adaptation current that hyperpolarizes the 
cell. Similar to $y$, $z$ quantifies the transport of ions but now through slow channels.

An interesting feature
of this model is that the dynamics associated to the $z$ variable are very slow when compared to those of
$x$ and $y$. The authors investigate three values for the current: $I=0.4$, $I=2$ and $I=4$. For $I=3.25$
a ``random'' (chaotic) burst structure has been reported. 
% rkHR84.m; dvHR84.m; simHR84RK.m

%----------------------------------------------------------------------------
\subsection{Izhikevich's spiking neuron Model}

\begin{figure}[ht]
  \centering
  \begin{tabular}{cc}
    \includegraphics[width=0.45\textwidth]{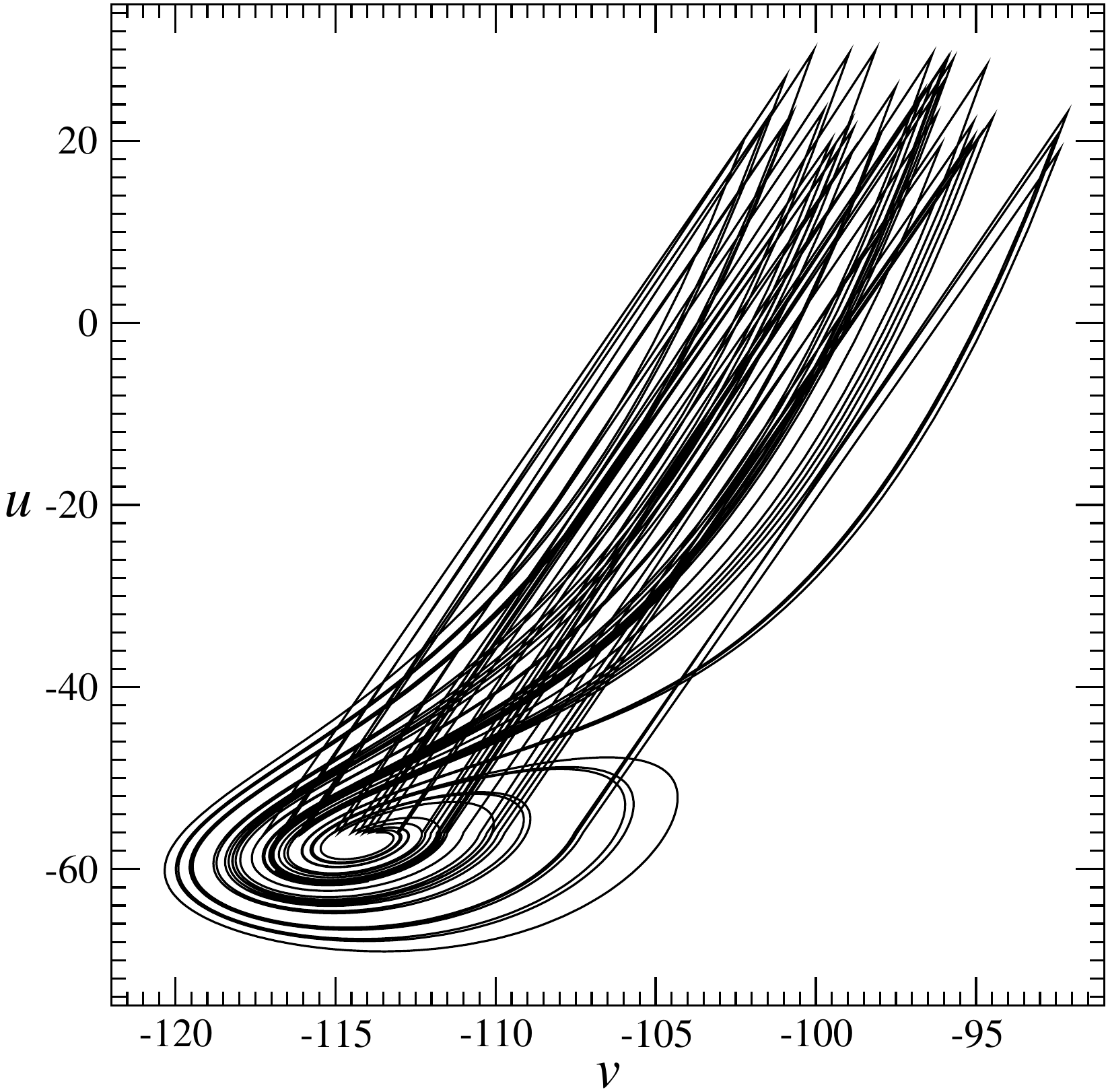} &
    \includegraphics[width=0.45\textwidth]{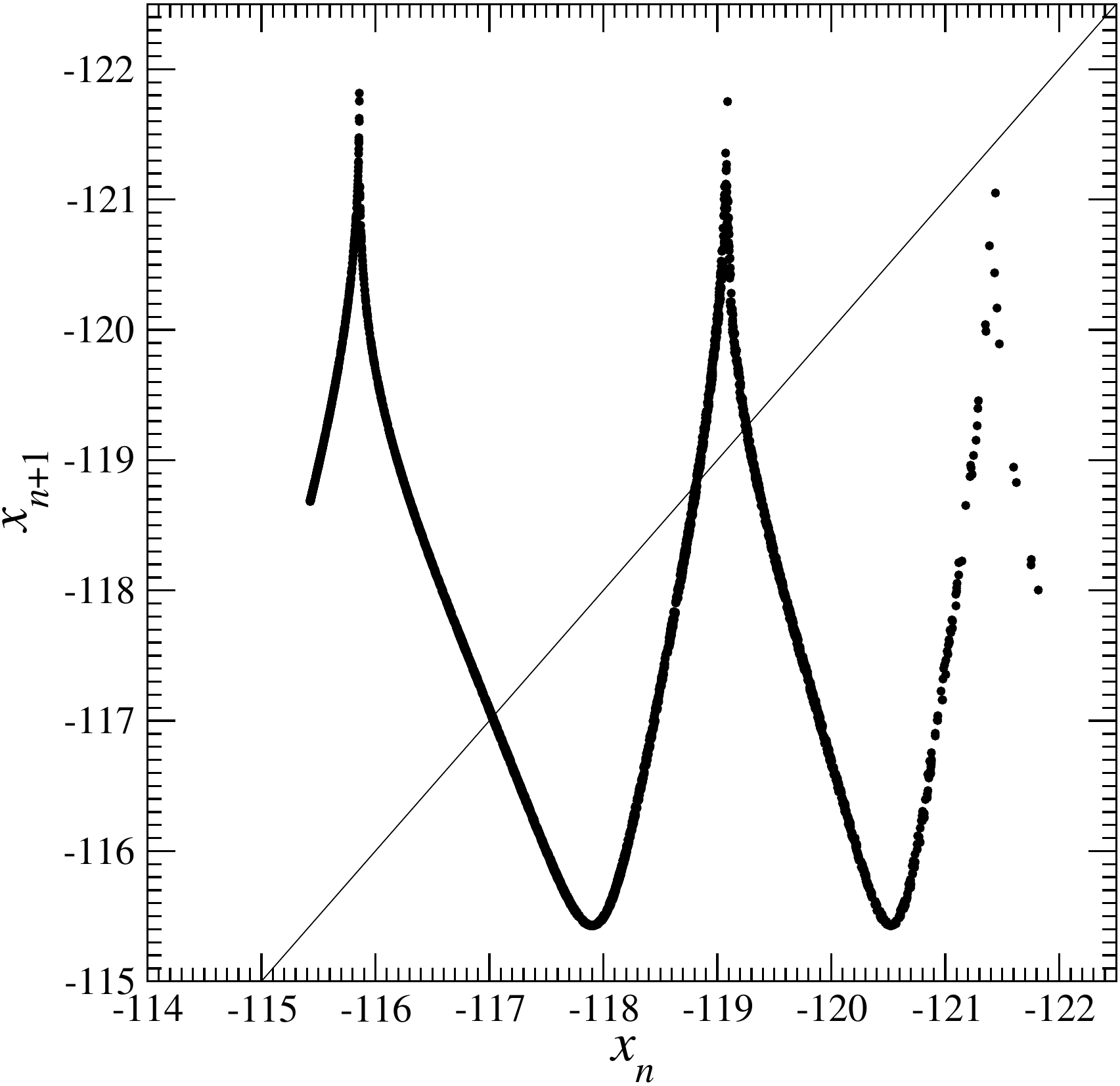} \\
    (a) Chaotic attractor & (b) First-return map \\[-0.2cm]
  \end{tabular}
  \caption{Chaotic behavior produced by the Izhikevich's model. Parameter 
values: $(a,\,b,\,c,\,d)=(0.2,\,2,\,-56,\,-16)$ with $I_{\rm syn}=-99$. The 
six-branches first-return map to a Poincar\'e section of (\ref{I03}) is typical
of a ``funnel'' chaotic behavior encountered in the R\"ossler system for 
$a \approx 0.540$, $b=2$ and $c=4$.  }
  \label{izhimap}
\end{figure}

A simple model that reproduces spiking and bursting behavior of known types of cortical neurons
was given by Eugene Izhikevich in 2003. It combines aspects of the Hodgkin-Huxley model and
of integrate-and-fire neuron models. The result is a surprisingly simple model that reproduces a
rich variety of dynamical regimes. The equations are \citep{izh/03}: 
\begin{eqnarray}
  \label{I03}
  \left\{ 
  \begin{array}{l}
    \dot{v} = 0.04v^2+5v+140-u+I_{\rm syn}  \\
    \dot{u} = a(bv-u)  ,
  \end{array} \right. \\
  {\rm if}~ v\ge 30~{\rm then} 
  \left|
    \begin{array}{l}
    v \leftarrow c \\
    u \leftarrow u+d \, ,
    \end{array}
  \right. \nonumber 
\end{eqnarray}

\noindent
where $v$ represents the membrane potential and $u$ represents the membrane 
recovery variable $(a,\,b,\,c,\,d)$ are constant parameters and $I_{\rm syn}$ 
accounts for the synaptic or injected dc currents. Depending on the parameter 
values, a rich variety of dynamical regimes are possible. For instance
$(a,\,b,\,c,\,d)=(0.02,\,0.2,\,-50,\,2)$ with $I_{\rm syn}=10$ results in 
chattering \citep{izh/03}  and $(a,\,b,\,c,\,d)=(0.2,\,2,\,-56,\,-16)$ with 
$I_{\rm syn}=-99$ results in chaotic
firing \citep{izh/04} as evidenced by the first-return map to a 
Poincar\'e section (Fig.\ \ref{izhimap}). The fact that a second-order model 
could produce chaos should not cause surprise because the switching 
function plays the role of the reinjection 
mechanism as governed by the third equation of the R\"ossler system for 
instance, variables $u$ and $v$ being associated with an oscillator as variable
$x$ and $y$ of the R\"ossler system. Izhihevich's model is therefore a 
``truncated'' model in the sense that all underlying mechanisms are not 
explicitly described. In a sense, the state of the switch acts as a ``hidden 
state variable''.
% rkI03.m; dvI03.m; simI03RK.m

%=======================================================
\section{Mathematical and Numerical Tools}
\label{tools}

In this section we briefly review the observability coefficients and 
multivariate singular spectrum analysis (M-SSA). Because of confusion in the 
literature as to the development of observability coefficients \citep{su_eal/17}, here a brief historical overview
is provided. 

The concepts of observability and controllability for linear systems are due to 
Rudolf Kalman \citep{kal/60}. These were extended to nonlinear systems over a decade
later as discussed in \citep{her_kre/77}. In both cases such concepts were ``yes'' or ``no'' concepts.
Bernard Friedland suggested computing a conditioning number of a symmetric matrix obtained
from the {\it linear}\, observability or controllability matrices as a way of getting a continuous
function of the parameters instead of a binary (yes or no) classification \citep{fri/75}.
In fact, it was argued that although a similarity transformation of coordinates would not
change the rank of the observability or controllability matrices, and therefore would not
alter the resulting classification, the indices proposed by Friedland are sensitive to
such a transformation \citep{agu/95}, to changes in parameters and, in the nonlinear case, to
the location in state space. The concept of a continously varying quantification
of observability was adapted to nonlinear dynamical systems in \citep{let_eal/98,let_agu/02}
where the jacobian matrix {\it of the vector field}\, was used in the analysis. Later on, it
was shown that the jacobian matrix {\it of the map}\, between the original and embedding
spaces coincided with the nonlinear observability matrix based on Lie derivatives \citep{let_eal/05pre}.
Hence, the quantification of observability was then performed using such a matrix. The 
extension to multivariate embeddings and the relation to Takens' theorem were presented
in \citep{agu_let/05}. The procedure proposed in \citep{let_eal/05pre} is briefly reviewed in
the next section.

%----------------------------------------------------------------------------
\subsection{Numerical observability coefficients}
\label{oc}

Consider the autonomous system $\dot{\mb{x}} = f(\mb{x})$, where $\mb{x}\in \bbR^n$ is the
state vector and $f: \bbR^n \mapsto \bbR^n$ is the vector field. Consider further the measurement
function $h: \bbR^n \mapsto \bbR$ such that $s(t)=h(\mb{x})$, where $s \in \bbR$ is referred to
as the observable or recorded variable. 
The case for which $s \in \bbR^p, p>1$ has been investigated and reported in \citep{agu_let/05}.
Differentiating $s(t)$ with respect to time yields
\begin{equation}
\label{c250105}
  \dot{s}(t) = \frac{d}{dt} h (\mb{x}) = \frac{\partial h}{\partial
  \mb{x} } \dot{\mb{x}}= \frac{\partial h}{\partial
  \mb{x} } \mb{f}(\mb{x})= {\cal L}_f h (\mb{x}) ,
\end{equation}
where
${\cal L}_f h (\mb{x})$ is the Lie derivative of $h$ along the vector
field $\mb{f}$ \citep{isi/95}.

The general observability matrix can be written  as \citep{her_kre/77}
\begin{equation}
\label{newdef}
  {\cal O}_s (\mb{x}) = 
  \left[
    \begin{array}{c}
      \displaystyle
      \frac{\partial {\cal L}^0_f h (\mb{x})}{\partial \mb{x}} \\[0.4cm] 
      \vdots \\[0.4cm]
      \displaystyle
      \frac{\partial {\cal L}^{n-1}_f h (\mb{x})}{\partial \mb{x}}
    \end{array}
  \right] ,
\end{equation}
where $s$ indicates that ${\cal O}_s (\mb{x})$ refers to the system observed 
from $s(t)$. In the case $h(\bx)$ returns only one of the state variables 
matrix (\ref{newdef}) can be rewritten as
\begin{equation}
\label{newdef2}
  {\cal O}_s (\mb{x}) = 
  \left[
    \begin{array}{c}
      \displaystyle
      C\\
C\tilde{A}\\
\vdots \\
C\tilde{A}^{n-1}
    \end{array}
  \right] ,
\end{equation}
\noindent
where $C=[1~0 \ldots 0]$ if $h(\bx)$ returns the first state variable, 
$C=[0~1~0 \ldots 0]$ if $h(\bx)$ returns the second, and so on. Also
\begin{equation}
  \label{Amatn}
  \tilde{A}^{j+1} =
  \left[
    \displaystyle \frac{\partial {\cal L}^j_{f} f_i(\mb{x}) }{\partial \mb{x}} 
  \right] ,~ i=1,2,\ldots, n
\end{equation}
for $~j=0,\ldots,n-2$, where 
\begin{equation}
\label{c270404}
  {\cal L}_f f_i(\mb{x}) =  \frac{\partial f_i (\mb{x})}{\partial
    \mb{x}} \mb{f}(\mb{x}) = \sum_{k=1}^n 
    \frac{\partial f_i (\mb{x})}{\partial \mb{x}} f_k 
\end{equation}
is the Lie derivative of the $i$th component of the vector field
$\mb{f}$ and the higher-order derivatives can be recursively
determined as
\begin{equation}
  {\cal L}^j_f {f_i}(\mb{x})  = {\cal L}_f
  \left[ \displaystyle {\cal L}^{j-1}_f {f_i}(\mb{x})  \right] ,
\end{equation}

\noindent
with ${\cal L}^0_f f_i (\mb{x}) = f_i (\mb{x})$. If ${\cal O}_s (\mb{x})$ is singular then there is no
global diffeomorphism between the original phase space and the $n$-dimensional space reconstructed
using $s$ and $n-1$ successive derivatives of it. Because the system is nonlinear, often ${\cal O}_s (\mb{x})$
may become singular or nearly singular at specific regions of state space at which the original dynamics
become poorly observable or nonobservable altogether.

Hence it is sometimes instructive to have an average measure of the numerical conditioning of ${\cal O}_s (\mb{x})$.
This is achieved by first computing along a trajectory $\mb{x}(t)$
\begin{equation}
  \label{do}
  \delta_s (\mb{x}) = \frac{\mid \lambda_{\rm min} 
   [{\cal O}_s(\mb{x})^T{\cal O}_s(\mb{x})]\mid}{\mid 
      \lambda_{\rm max} [{\cal O}_s(\mb{x})^T{\cal O}_s(\mb{x})] \mid} ,
\end{equation}
where $\lambda_{\rm max}[ \cdot ]$ indicates
the maximum eigenvalue of the argument estimated at
point $\mb{x}(t)$ (likewise for $\lambda_{\rm min}$); $0 \le
\delta (\mb{x}) \le 1$, and the lower bound is reached when the system
is not observable at point $\mb{x}$. 
Finally, averaging $\delta_s(\mb{x})$ along a trajectory over the interval $t\in[0;~T]$ yields
\begin{equation}
\label{do2}
  \delta_s = \frac{1}{T} \sum_{t=0}^{T} \delta_s({\mb{x}(t)}),
\end{equation}
where $T$ is the final time considered and, without loss of generality,
the initial time was set to be $t=0$. 
%

%----------------------------------------------------------------------------
\subsection{Symbolic Observability Coefficients}
\label{so}

The advantage of the numerical observability coefficients is that they take 
into account the domain of the state space actually visited by the trajectory 
and, consequently, whether the neighorhood of the singular observability 
manifold is visited or not. Nevertheless, these observability coefficient are
not normalized and cannot be used to compare different dynamical systems. To
overcome such a problem, symbolic observability coefficients were introduced
\citep{let_agu/09pre}. The underlying idea is that the more complicated the 
determinant det~$\mathcal{O}_s$ of the observability matrix, the less observable.
Although the analytical computation of det~$\mathcal{O}_s$ can be
a nearly impossible task for a five-dimensional rational system, 
the complexity of det~$\mathcal{O}_s$ can be assessed simply by counting the number of 
linear, nonlinear and rational terms in it, without paying attention
to its exact form \citep{Bia15}. This is computed from the Jacobian matrix of 
the system which is transformed into symbolic form, using 1, $\1$, and 
$\2$ for linear, nonlinear and rational elements, respectively. The 
observability matrix is then constructed using symbolic algebra detailed in
\citep{Bia15}. The symbolic observability coefficient is thus defined as
\begin{equation}
  \begin{array}{rl}
    \eta_{s^n} =  & \displaystyle
    \frac{N_1}{N_1 + N_{\1} + N_{\2}} %\\[0.4cm]
%    & \displaystyle
    + \frac{N_{\1}}{\left( \displaystyle \mbox{max}(N_1,1) + N_{\1} 
          + N_{\2} \right)^2} \\[0.4cm]
    & \displaystyle
    + \frac{N_{\2}}{\left( \displaystyle \mbox{max}(N_1,1) + N_{\1} 
          + N_{\2} \right)^3} ,
  \end{array}
\end{equation}

\noindent
where $N_1$, $N_{\1}$ and $N_{\2}$ are the numbers of symbolic terms 1, $\1$ 
and $\2$, respectively. These symbolic coefficients are very promising for 
assessing the observability of large systems and networks 
\citep{Let17}.
According to \citep{sen_eal/16}, the observability can be considered ``good'' 
when $\eta_{s^3} > 0.75$, meaning that most likely the determinant of the 
observability matrix is linear if not constant; consequently the influence of
the singular observability matrix is not very important.

%----------------------------------------------------------------------------
\subsection{Observability coefficients from data}
\label{ocd} 

The procedures reviewed in Sec.\,\ref{oc} and Sec.\,\ref{so} require the knowledge of the system equations.
Motivated by the fact that in practice equations are not always available, an 
alternative procedure was proposed in \citep{agu_let/11}. However, 
observability is, by definition, related to the equations of the vector field. 
Hence estimating coefficients from data is only an indirect way of assessing 
observability from some of its {\it signatures}\, found in a reconstructed
(embedding) space, as explained next.

The rationale behind the method in \citep{agu_let/11} is that in the embedding space of a
system with poor observability conveyed by a recorded time series, trajectories are either pleated or 
squeezed. Such features result in a more complex {\it local}\, structure in the embedding space.
On the other hand, in the space reconstructed using good observables, very often, trajectories
are unfolded comfortably and that translates into a more simple local structure of such a
space. The SVDO coefficients hence quantify, using the singular value decomposition (SVD) of
a trajectory matrix, the local complexity of the reconstructed space. Simpler structures are associated
with better observability whereas more complex local structures with poorer observability. A key
point to be noticed here is that SVDO cannot quantify observability {\it per 
se}, which by definition
would require the vector field equations, but rather are indicators of the {\it average local complexity}\,
of a reconstructed space, which often -- but not always, as will be seen shortly -- correlates
with observability.

This procedure assumes that the embedding space is reasonable, in the sense that the dimension
and delay time -- in the case of time delay coordinates -- have been correctly chosen. Hence
if trajectories are either pleated or squeezed this is attributed to poor observability rather
than to a bad choice of embedding parameters. In this paper we use time delay coordinates.
Although the results will be reported for a given time delay and embedding dimension, numerical
studies with different values of such parameters show no change in the ranking of the variables
in terms of observability.

%----------------------------------------------------------------------------
\subsection{Multivariate singular spectrum analysis}
\label{mssa}

Here, the use of the structured-varimax multivariate spectrum analysis 
(svM-SSA) for phase synchronization phenomena, originally proposed by Groth 
and Ghill \citep{gro_ghi/11}, is briefly reviewed (for a full discussion of the 
method, along with the introduction of a Monte Carlo based statistical 
confidence test, see \citep{gro/15}). Recently, it was argued that the 
explanatory power of svM-SSA on the mechanism of phase synchronization is 
greatly enhanced by the use of a single state variable, as compared to the 
original approach using all of them, as long as this variable provides good 
observability of the dynamics as compared to the rest \citep{por_agu/16chaos}.

\begin{figure}\centering
\includegraphics[width=0.7\linewidth]{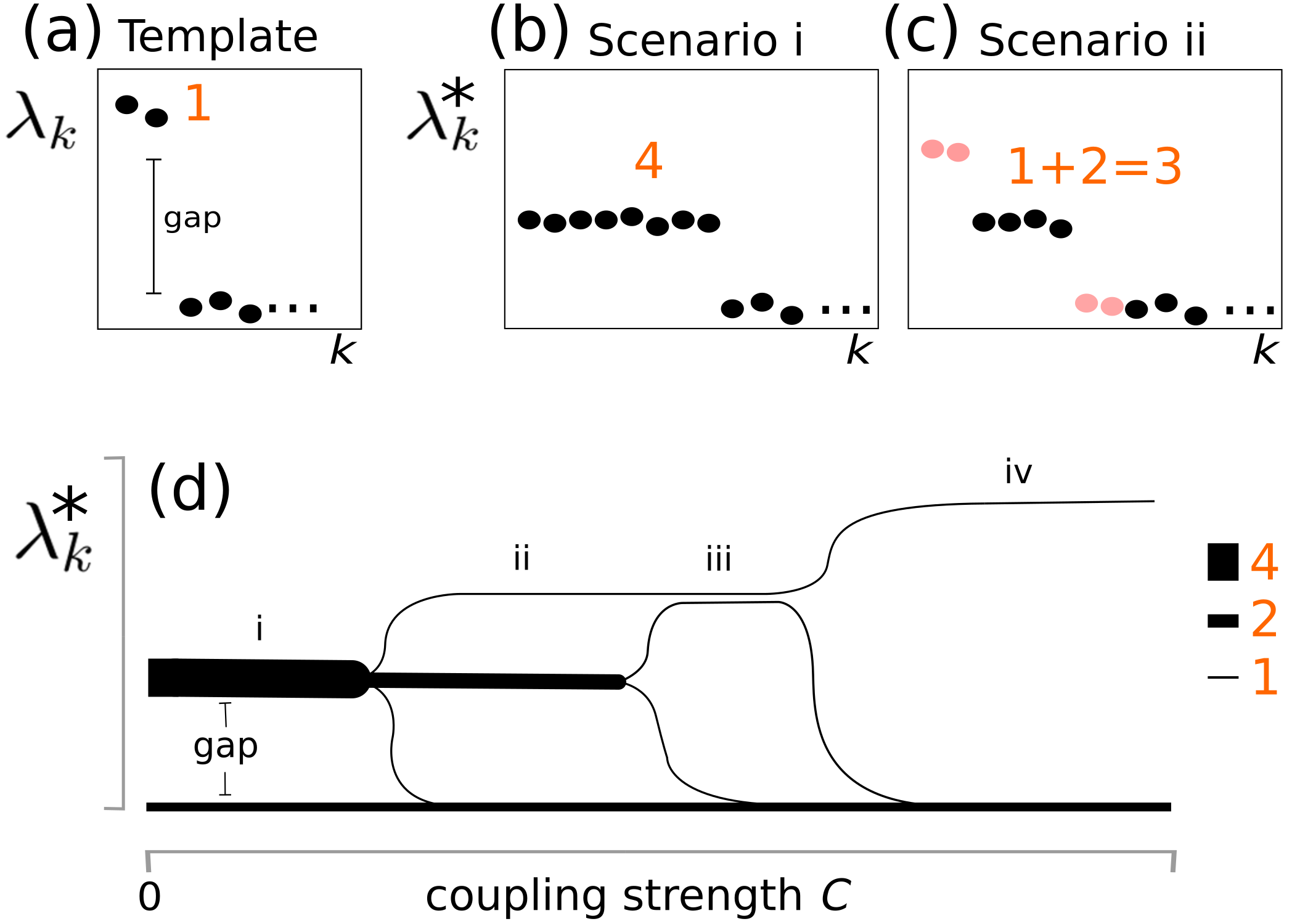}
\caption{\label{fig.mssa.tutorial}Schematic representation of an svM-SSA for $J=4$ idealized coupled oscillators. 
(a)~The {\it template}, shows the signature of a single oscillatory mode, identified by a unique ($\mu=1$) pair of singular 
values $\lambda_k$ separated from the ``noise floor'' by a clear gap. (b)~When the oscillators are not phase 
synchronized, the svM-SSA shows $\mu=4$ pairs of {\it modified variances} $\lambda_k^*$  (similar to ``four 
concatenated templates'') before a small gap, indicating four distinct oscillatory modes. (c)~When two oscillators phase synchronize (PS), 
the corresponding {\it single}\,
$\lambda_k$ pair is larger than the other two associated with the remaining non-synchronized oscillators, hence
$\mu=3$ (before the gap). Note that the increase of $\lambda_k^*$ for one pair due to the phase synchronization of two oscillators
is accompanied by another pair merging with the noise floor (both pairs are indicated in orange). (d) $\lambda_k^*$ 
for an increasing coupling strength $C$: (i)~oscillators start non-synchronized -- like in (b); (ii)~two oscillators forms a PS 
cluster when a pair of $\lambda_k^*$ increase and another pair fall to the noise floor -- as the orange pairs in (c); (iii)~other two 
oscillators form a second PS cluster, and again there is an increase in a $\lambda_k^*$ pair while the other merges with the 
noise floor (and $\mu=2$); (iv) finally the two clusters merge into a single PS one. The number $\mu$ of $\lambda_k^*$ 
pairs above the noise floor (equivalently, of different oscillatory modes detected) is indicated by the thickness of the lines.}
\end{figure}

Consider $J$ coupled oscillators and the respective time series of length $N$ 
of the variable that provides the best observability. Each time series is 
split in windows of $m$-data points \citep{bro_kin/86}, and a full augmented ``trajectory matrix'' ${\bf X}=[{\bf X}_1, ..., {\bf X}_J]\in\bbR^{N-m+1, Jm}$ is formed by concatenating the individual time series. In the M-SSA literature, the parameter $m$ is called window width or embedding dimension: we prefer to use the former term. The svM-SSA starts by the eigendecomposition of the covariance matrix ${\bf C}={\bf X}^T{\bf X}/(N-m+1)$ as ${\bf \Lambda}={\bf E}^T{\bf C}{\bf E}$. To overcome a known mixture of the eigenvectors related to the individual subsystems, a structured-varimax rotation \citep{gro_ghi/11,por_agu/16pre} is performed on the first $S$ eigenvectors, ${\bf E}_S^*={\bf E_S}{\bf T}$. 
Finally, the modified variances $\{\lambda_k^*\}_{k=1}^S\equiv {\rm diag}({\bf \Lambda}_S^*)$ are obtained through ${\bf \Lambda}_S^*={\bf T}^T{\bf \Lambda}_S{\bf T}$ and they encode information about the underlying structure of the data: a single high value is related to a trend; pairs of nearly equal values reflect oscillatory modes; near zero values are associated with noncoherent oscillations and will be referred to as the noise floor.
Hence it is possible to infer about phase synchronization,  {\it without any prior definition}\, of how to estimate the 
oscillators phases, from the evolution of $\lambda_k^*$ pairs associated to the oscillatory modes in the data, as  
schematically illustrated in Fig.~\ref{fig.mssa.tutorial} for $J=4$ idealized oscillators.

%

%=======================================================
\section{Results on Observability}
\label{nr}

Here we provide numerical results about the observability of the models in 
Section \ref{nm} which were integrated using a 4th-order Runge-Kutta algorithm 
with integration step $h=0.01$.

%----------------------------------------------------------------------------
\subsection{Hodgkin-Huxley Model}

Given the complexity and dimension of the Hodgkin-Huxley model, the observability matrices are too
large to be shown here. Also, since (\ref{HH52}) is a biophysically-based model, when investigating
observability properties one should keep in mind what variables are actually recordable. Out of
the four state variables of this model only the membrane potential $V$ is recordable, the other variables
being dimensionless quantites. Nonetheless, for the sake of completion, we here report the observability
coefficients for the four state variables:
$\delta_V= 1.1122 \cdot 10^{-7}$, $\delta_n=1.0872 \cdot 10^{-6}$
$\delta_m=5.1427 \cdot 10^{-9}$ and $\delta_h=4.8986 \cdot 10^{-6}$.
% obsHH52.m; obsHH52_sym.m;
The symbolic Jacobian matrix is
\begin{equation}
  {\cal J}^{\rm sym} =
  \left[
    \begin{array}{cccc}
      \1 & \1 & \1 & \1 \\
      \2 & \1 & 0 & 0 \\
      \2 & 0 & \1 & 0 \\
      \2 & 0 & 0 & \1 
   \end{array}
  \right]
\end{equation}

\noindent
from which the symbolic observability coefficients $\eta_{V^4} = 0.12$, 
and $\eta_{n^4} = \eta_{m^4} = \eta_{h^4} =  0.19$ can be obtained.

Because the state variables $n$, $m$ and $h$ cannot be measured, a different
procedure was followed that is made available by using the indirect assessment of observability from data proposed in \citep{agu_let/11}.
Hence  the membrane potential $V$ and the currents $I_{\rm K}$, $I_{\rm Na}$ and $I_l$ were considered
as candidate variables to be used in reconstructing a phase space for the dynamics. 

In so proceding, the following 
SVDO coefficients were found $S_V=0.109\pm0.0009$, 
$S_{I_{\rm K}}=0.093\pm0.0015$, $S_{I_{\rm Na}}=0.054\pm0.0024$, $S_{I_l}=0.167\pm0.0028$  
for $I=-10$ which show that the best variables for reconstructing a phase space using delay coordinates are
$V$ and ${I_l}$. From a practical point of view, the best variable to be recorded is probably the potential $V$ as 
$I_l$ is the ionic current of {\it all other}\, ions besides those of potassium and sodium.
The reported values were computed using a 5-dimensional embedding space and a common delay time ($\tau=100$ sampling intervals)
for the three variables. The reported values are mean plus-minus one standard deviation over 10 Monte Carlo runs.
%$S_n=2.227\pm0.0127$, %$S_m=0.165\pm0.0040$, $S_h=2.399\pm0.0324$  (not used)
% obsHH52_svd_delay.m

%----------------------------------------------------------------------------
\subsection{FitzHugh-Nagumo Model}
\label{obsFHN}

The observability matrix for model (\ref{FN61}) when $x$ is recorded is:
% obsFN61_sym.m
%
\begin{eqnarray}
{\cal O}_x = \left[
\begin{array}{ccc}
1 & 0 \\
-c(x^2-1) & c  \\
\end{array} \right] ,
\end{eqnarray}

\noindent
with determinant det$({\cal O}_x)=c$, hence unless $c=0$ the system is observable from the $x$
variable, although observability could be poor for very small values of $c$. Recording the recovery
variable $y$ yields the observability matrix:
% obsFN61_sym.m
%
\begin{eqnarray}
{\cal O}_y = \left[
\begin{array}{ccc}
0 & 1 \\
-1/c & -b/c  \\
\end{array} \right] ,
\end{eqnarray}

\noindent
also with constant determinant det$({\cal O}_y)=1/c$. Hence for very high values of $c$, the $y$ 
variable conveys worse observability of the dynamics. 

\begin{figure}
		\centering
		\includegraphics[width=0.7\textwidth]{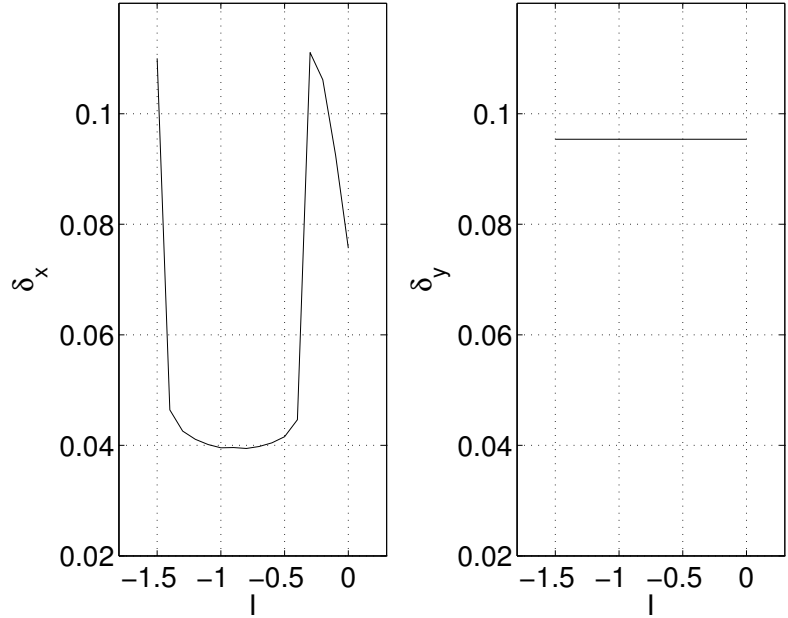}
		\caption{\label{FNdeltas} Observability coefficients computed for the FitzHugh-Nagumo model (\ref{FN61}).
		with $(a,\,b,\,c)=(0.7,\,0.8,\,3)$ and the current $I$ was maintained constant
		for each value in the range $-1.5 \le I \le -0.4$.}
		% obsFN61.m
\end{figure}

Figure~\ref{FNdeltas} shows the observability coefficients for the FitzHugh-Nagumo model. Notice that
because ${\cal O}_x$ depends on variable $x$ it varies along the limit cylce on the $x$-$y$ plane which,
in turn is affected by $I$. Contrary to this ${\cal O}_y$ is constant throughout the phase plane and is not
influenced by the stimulus. Both observability coefficients $\delta_x$ and $\delta_y$ are of the same order
of magnitude for the chosen parameters.
Hence, unless $c$ is very large or very small, both state variables are comparable in what concerns
observability, although measuring $y$ ensures more uniform performance along the limit cycle.
The symbolic observability matrix is
\begin{equation}
  \mathcal{O}_x^{\rm sym} =
  \left[
    \begin{array}{cc}
      1 & 0 \\
      \1 & 1
    \end{array}
  \right]
\end{equation}

\noindent
and the corresponding determinant is Det~$\mathcal{O}_x^{\rm sym} = 1 \otimes 
1$ where $\otimes$ is the multiplicative law between the symbols as
defined in \citep{Bia15}; only constant terms are involved in the symbolic
determinant and
the corresponding symbolic observability coefficient is therefore
$\eta_{x^2} = 1$. As long as $c$ is sufficiently different from 0, the symbolic 
observability coefficient does not overestimate the observability of
the system. With a similar approach, we found $\eta_{y^2} = 1$.

The SVDO for the Fitzhugh-Nagumo model are $S_x=0.301\pm0.0199$, $S_y=0.206\pm0.0058$ 
for $I=-0.4$ which confirm that the observability of both variables are comparable and that the
embedding space reconstructed with $y$ is rather more homogeneous.
These values were computed using a 3-dimensional embedding space and a common delay time ($\tau=100$ sampling intervals)
for the three variables. The reported values are mean plus-minus one standard deviation over 10 Monte Carlo runs.
 % obsFN61_svd_delay.m

%----------------------------------------------------------------------------
\subsection{Hindmarsh-Rose Model}
\label{obsHR}

The observability of the HR model has been considered recently in a pair of 
papers \citep{por_agu/16chaos,sen_eal/16}. In the first paper a modified 
version of (\ref{HR84}) was considered with linearly transformed coupled 
equations \citep{bel_eal/05prl}, and in the second symbolic observability 
coefficients \citep{let_agu/09pre} were computed. So here we compute the 
observability coefficients as used in \citep{let_eal/05pre} for model 
(\ref{HR84}). 
Following \citep{sen_eal/16} we use $(a,\,b,\,c,\,d)=(1,\,3,\,1,\,5)$ and 
$(r,\,s,\,x_1,\,I)=(0.001,\,4,\,-\frac{1+\sqrt{5}}{2},\,3.318)$.

The observability matrices for model (\ref{HR84}) are:
% obsHR84_sym.m
%
\begin{eqnarray}
{\cal O}_x = \left[
\begin{array}{ccc}
1 & 0 & 0\\
2bx - 3ax^2 & 1 & -1 \\
O_{31}^x & - 3ax^2 + 2bx - 1 & 3ax^2 - 2bx + r \\
\end{array} \right] ,
\end{eqnarray}

\noindent
where $O_{31}^x=(2bx - 3ax^2)^2 - rs - 2dx + (2b - 6ax)(-ax^3 + bx^2 + I + y - z)$. ${\cal O}_x$ becomes
singular for $r=1$ because in that case the two last columns become linearly dependent (LD), in fact, det$({\cal O}_x)=r-1$. Singularity
is not expected to happen given the range of values usually used for $r$, although a negative value for the determinant
may indicate that the original and reconstructed spaces are topologically equivalent, but {\it not}\,  orbitally equivalent, that is
the relative direction of trajectories may be different in each space;
\begin{eqnarray}
{\cal O}_y = \left[
\begin{array}{ccc}
0 & 1 & 0\\
-2dx & -1 & 0\\
O_{31}^y  & 1 - 2dx & 2dx\\
\end{array} \right] ,
\end{eqnarray}

\noindent
where $O_{31}^y=- 2d(-x -ax^3 + bx^2 + I + y - z) - 2dx(2bx - 3ax^2)$. ${\cal O}_y$ becomes
singular at $x=0$, as the last column becomes null, in fact det$({\cal O}_y)=4d^2x^2=100x^2$, for 
the parameters used. Although this situation could happen several times
during firing, the associated dynamics are so fast that the time the system spends close to $x=0$
is so short that this does not pose practical observability problems \citep{fru_eal/12}. Finally, 
the observability matrix when $z$ is recorded is:
\begin{eqnarray}
{\cal O}_z = \left[
\begin{array}{ccc}
0 & 0 & 1\\
rs & 0 & -r\\
rs(2bx - 3ax^2) - r^2s & rs & r(r-s)\\
\end{array} \right] ,
\end{eqnarray}

\noindent
which becomes singular at $rs=0$. In this case $\rho[{\cal O}_z]=1$, where $\rho[]$ stands
for the rank. This, added to the fact that $r$ is already quite small, shows that $z$ conveys poor
observability of the system. It is interesting to notice that although det$({\cal O}_z)=r^2s^2=1.6 \cdot 10^{-5}$
is constant this does not imply good observability.

\begin{figure}
  \centering
  \includegraphics[width=0.7\textwidth]{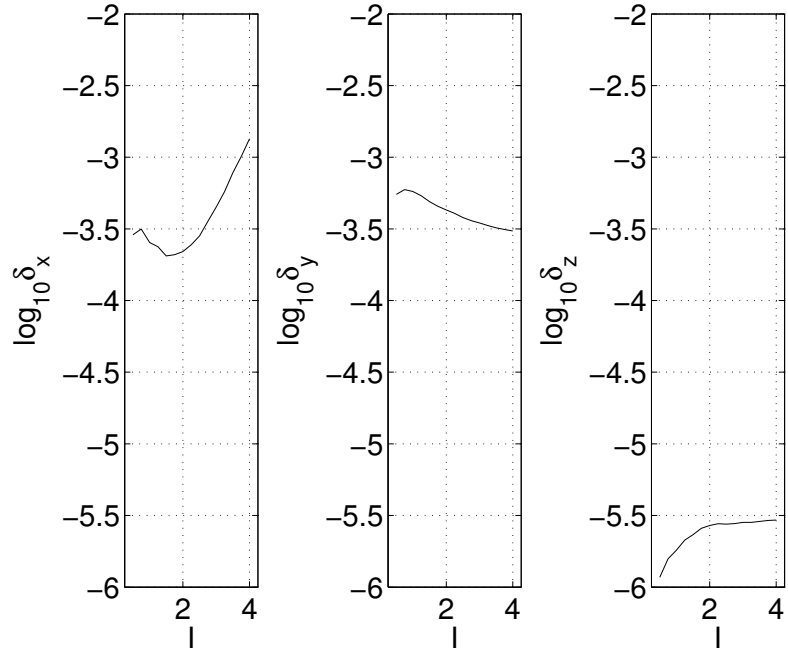}
  \caption{\label{HRdeltas}Observability coefficients computed for the 
Hindmarsh-Rose model. $(a,\,b,\,c,\,d)=(1,\,3,\,1,\,5)$, $(r,\,s,\,x_1)
=(0.001,\,4,\,-\frac{1+\sqrt{5}}{2})$ and the current was varied within the 
range $0.5 \le I \le 4$. In this plot the logarithm of the coefficients are 
shown.}
	% obsHR84.m
\end{figure}

In order to quantify observability, coefficients were computed and the results shown in Figure~\ref{HRdeltas}.
As seen, the fast variable provides much better observability than the slow variable $z$. Values of $I$ around 2
slightly favors $y$ compared to $x$, but as the current increases $x$ becomes slightly better. This difference
might not be critical in practice.

The symbolic observability matrix reads
\begin{equation}
  \mathcal{O}_x^{\rm sym} = 
  \left[
    \begin{array}{ccc}
      1 & 0 & 0 \\
      \1 & 1 & 1 \\
      \1 & \1 & \1 
    \end{array}
  \right]
\end{equation}

\noindent
and $\eta_{x^3} = 0.25$. Nevertheless, det~$\mathcal{O}_x = r-1$ and is clearly
not dependent on the location in the state space. It is never singular unless
$r \neq 1$. A global diffeomorphism could be therefore expected, that is, a 
full observability should be provided by this variable although
it could be very ill-conditioned, and therefore poor. This is one of the rare
cases where two nonlinear terms in the computation of the determinant are 
cancelling each other; the symbolic computation is therefore very different 
from an analytical computation. Note that as the dimension of the system increases, this
situation becomes less likely.
It has been argued that from the symbolic point of view, the 
coefficient $\eta_{x^3}$ should be corrected to be equal to 1 \citep{sen_eal/16}.  
As long as $r$ is significantly different from 1, taking $\eta_{x^3} = 1$ 
should be a fair estimation of the observability. From the two other 
symbolic observability matrix, we obtained $\eta_{y^3} = 0.56$ and $\eta_{z^3}
=1$.

\begin{figure}[h]
  \centering
  \begin{tabular}{ccc}
    \includegraphics[width=0.30\textwidth]{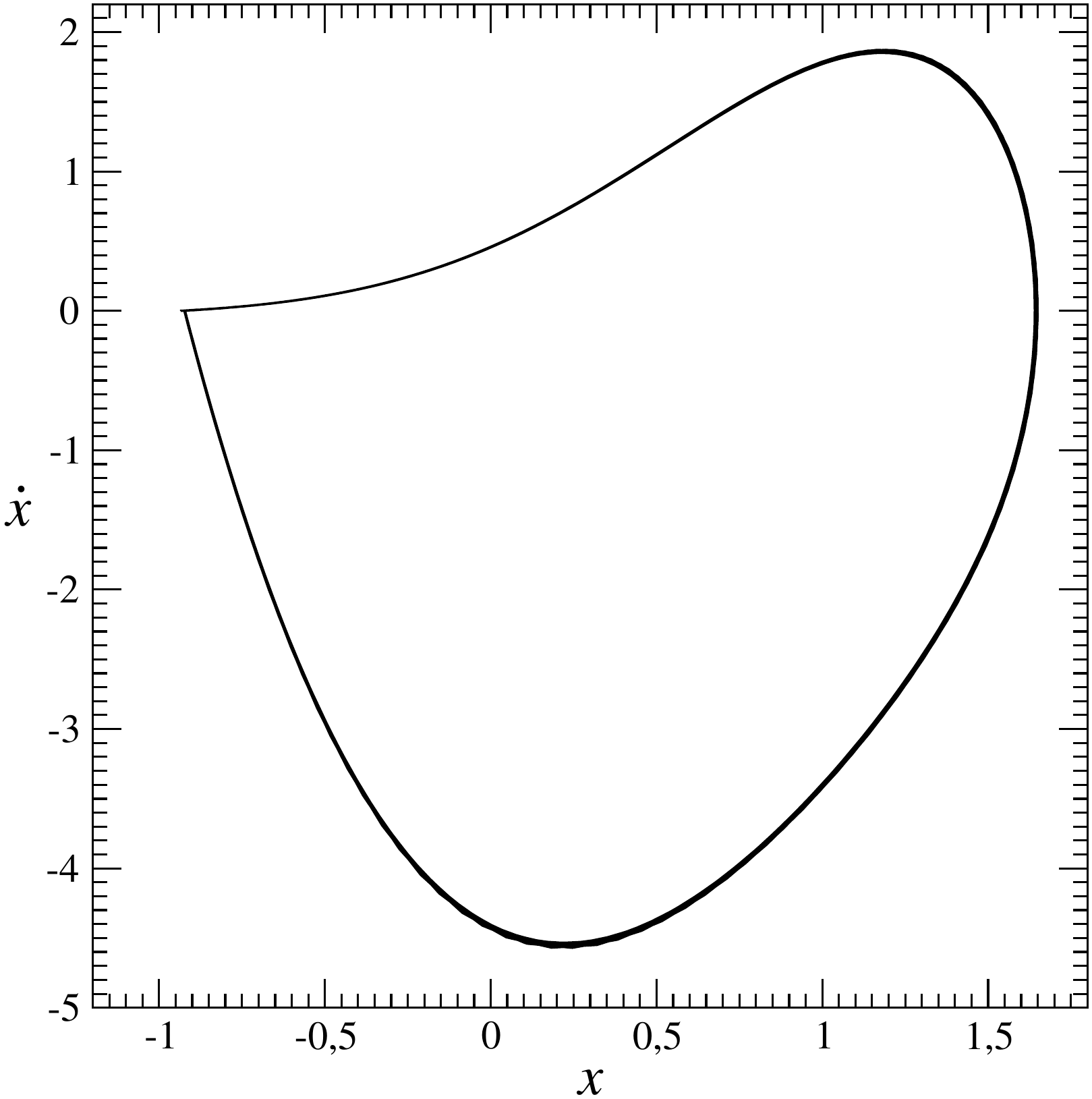} &
    \includegraphics[width=0.30\textwidth]{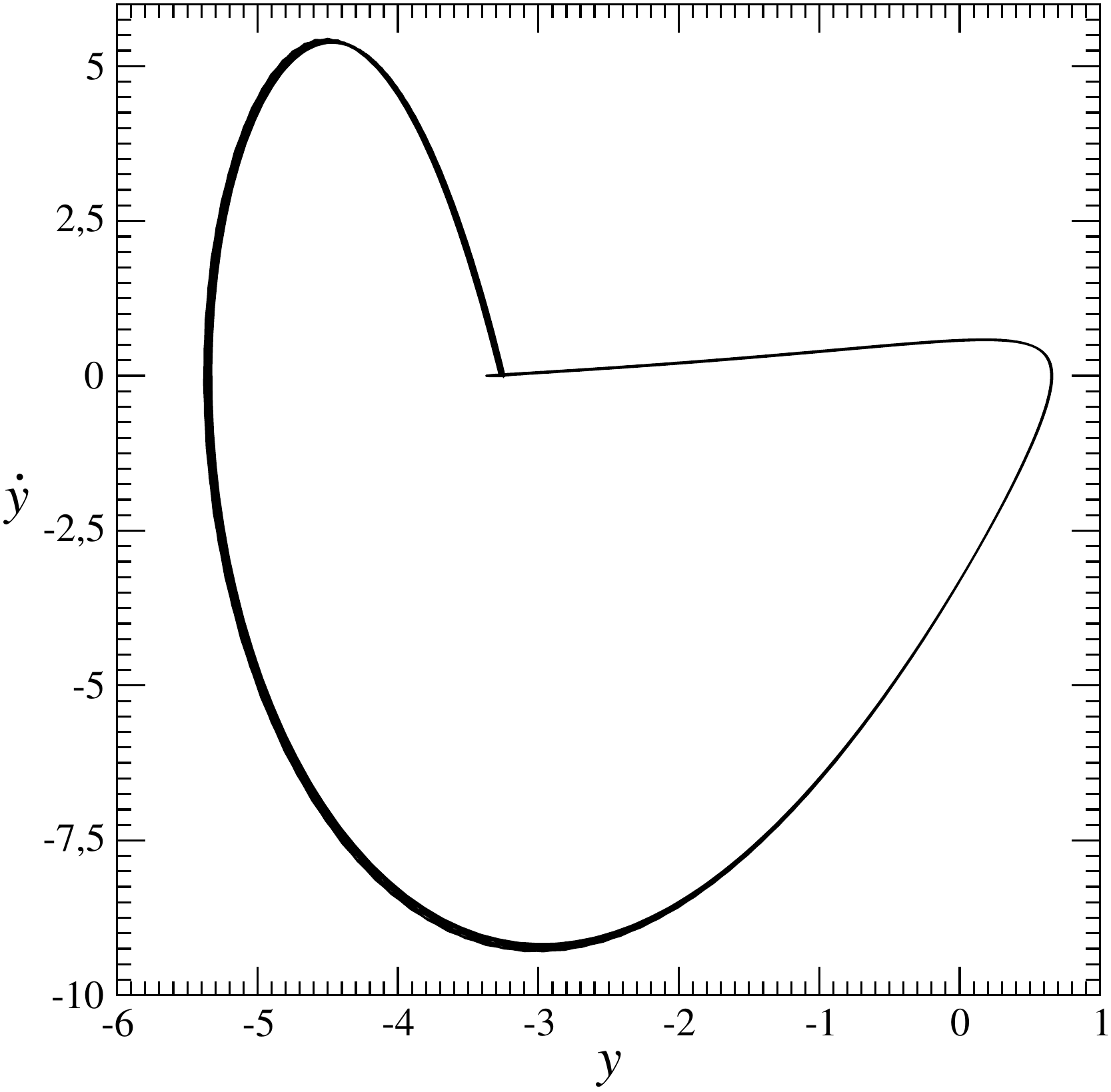} &
    \includegraphics[width=0.30\textwidth]{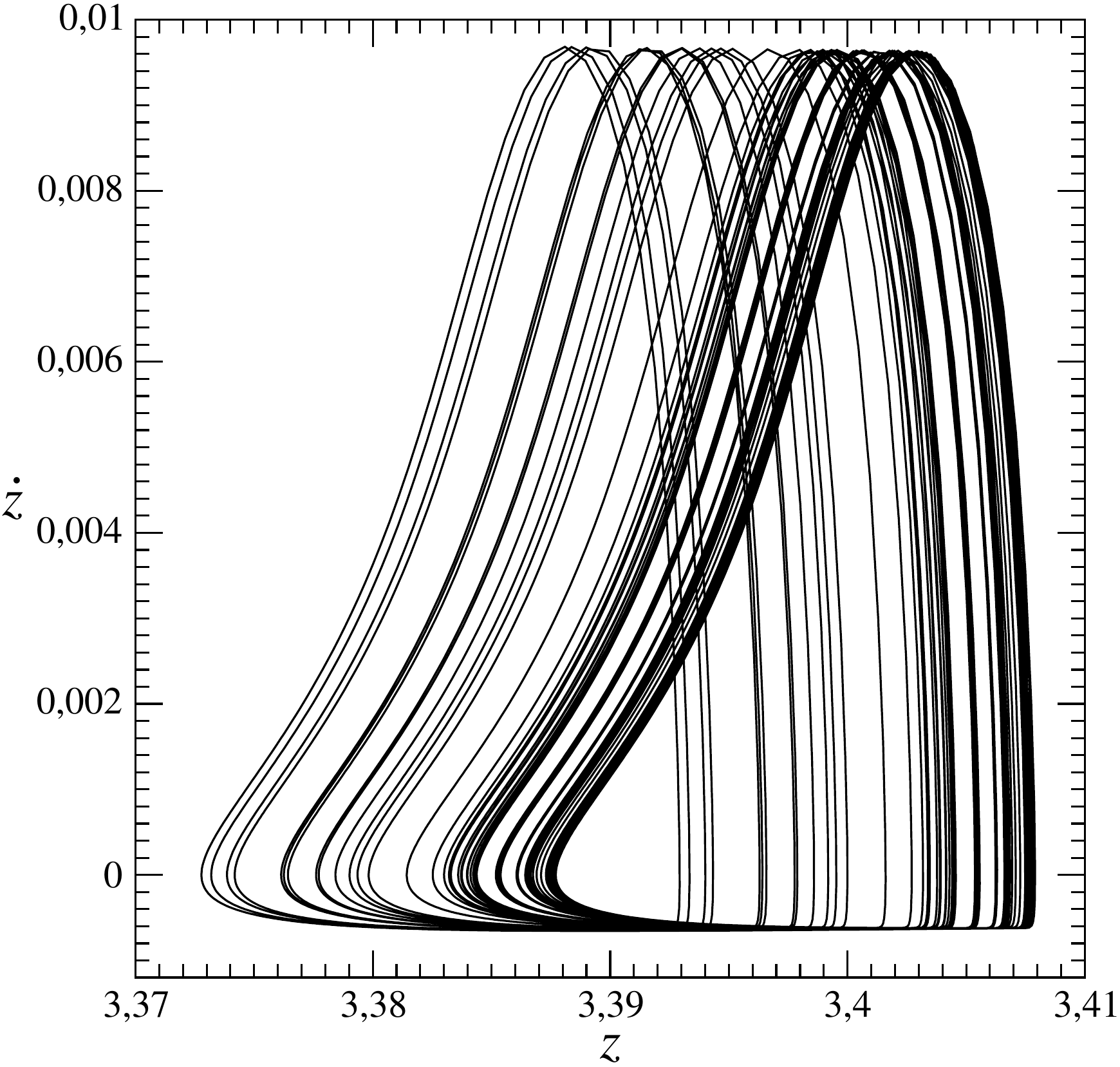} \\
    (a) $x$-$\dot{x}$ plane &
    (b) $y$-$\dot{y}$ plane &
    (c) $z$-$\dot{z}$ plane \\[-0.2cm]
  \end{tabular}
  \caption{Plane projection of the differential embedding induced by each 
variable of the Hindmarsh-Rose system.}
  \label{Hinder}
\end{figure}

The symbolic observability coefficient $\eta_{x^3}$ --- if not corrected ---
would suggest a rather poor observability of the dynamics underlying the 
Hindmarsh-Rose system from variable $x$. In fact, and contrary to what is 
provided from the determinant point of view, variable $x$ does not provide a 
good observability of the underlying dynamics. This is confirmed by a 
differential embedding induced by variable $x$ of the attractor produced by the 
Hindmarsh-Rose system (Fig.\ \ref{Hinder}a) which clearly shows that the 
chaotic nature of the behavior is poorly evidenced, contrary to what is 
observed when the differential embedding induced by variable $z$ is used
(Fig.\ \ref{Hinder}c). Contrary to what was recommended in \citep{sen_eal/16}, 
the symbolic observability coefficient $\eta_{x^3}$ should {\it not}\, be changed
since it correctly reveals that observability of the dynamics is, in fact, poor.

The SVDO for this system are $S_x=0.665\pm0.1557$, $S_y=0.449\pm0.1566$ and $S_z=186.628\pm33.6825$,
for $I=3.318$ and $S_x=0.416\pm0.0368$, $S_y=0.201\pm0.0282$ and $S_z=19.452\pm0.6102$ for $I=2$
(mean plus-minus one standard deviation over 10 Monte Carlo runs).
These values were computed using a 4th dimensional embedding space and a common delay time ($\tau=100$ sampling intervals)
for the three variables. Numerical experimentation with other values did not change the ranking of the variables.
% obsHR84_svd_delay.m

For both values of $I$ we find that $x$ and $y$ variables have similar features 
(as for $\delta_x$ and $\delta_y$), but the great difference is that $S_z$ 
suggests that $z$ provides much better ``observability''. 
This was also the case of the symbolic observability coefficients (if
not corrected). The reconstructed attractors --- using delay or derivative 
coordinates ---, evidence that the chaotic nature of the behavior produced 
by the Hindmarsh-Rose model is better evidenced by variable $z$. The values of the observability coefficient 
$S_s$ also confirm these results. This does not necessarily mean that $z$ is a
good observable. In fact, the fast dynamics (spikes) are practically invisible 
from $z$ --- which corresponds to the flat bottom of the attractor in 
Figure~\ref{Hinder}c. In few words it seems that $z$ is the best observable for 
slow dyamics (and chaos) whereas $x$ and $y$ convey information on the spikes, 
in the chattering regime.

%----------------------------------------------------------------------------
\subsection{Izhikevich's spiking neuron Model}
\label{obsIZ03}

The observability matrix for model (\ref{I03}) when $v$ is recorded is:
% obsI03_sym.m
%
\begin{eqnarray}
  {\cal O}_v = \left[
  \begin{array}{ccc}
    1 & 0 \\
    (2v/25)+5 & -1  \\
  \end{array} \right] ,
\end{eqnarray}
\noindent
with determinant det$({\cal O}_v)=-1$. Recording the recovery
variable $u$ yields:
% obsI03_sym.m
%
\begin{eqnarray}
  {\cal O}_u = \left[
  \begin{array}{ccc}
    0 & 1 \\
    ab & -a  \\
  \end{array} \right] ,
\end{eqnarray}

\noindent
also with constant determinant det$({\cal O}_u)=-ab$. Because ${\cal O}_u$ is constant,
the observability features do not vary in phase space, unlike for the case when the
membrane potential $v$ is measured. In particular, switching $v$ will directly affect
${\cal O}_v$. A peculiar aspect of this model is that both determinants are negative.

$\delta_v=0.3280$ and $\delta_u=1.5987 \cdot 10^{-5}$ for $(a,\,b,\,c,\,d)=(0.02,\,0.2,\,-50,\,2)$ with $I=10$,
and $\delta_v=0.3948$ and $\delta_u=0.1459$ for $(a,\,b,\,c,\,d)=(0.2,\,2,\,-56,\,-16)$ with $I=-99$. 
Hence in both scenarios the membrane potential is the variable that provides best observability. More interestingly,
the recovery varible is very poor observability-wise in the chattering regime, whereas it is comparable to the
membrane potential in the chaotic regime.
By definition, the symbolic observability coefficients are not dependent on the
parameter values. We thus found 
$\eta_{u^2} = \eta_{v^2} = 1$ (for chattering and chaotic regimes)
which would suggest that any of these two variables offers a full 
observability.
It must be clear that by construction this is an 
approximation to observability since the switching 
mechanism is not fully described in terms of differential equation (at least 
a third variable would be necessary for this) and, consequently, there is no
available technique so far to rigorously assess the observability of such a system.

The SVDO coeficients ($d_{\rm e}=3$ and $\tau=100$ sampling intervals) for model (\ref{I03}) are 
$S_v=0.858\pm0.2721$, $S_u=0.173\pm0.1164$ for chattering and 
$S_v=0.750\pm0.2869$, $S_u=8.376\pm0.5058$ for the chaotic regime. It is worth mentioning that
in both analyzes, the observability using the recovery variable improves when the dynamics are
chaotic. It should be pointed out that the discontinuities of the trajectories
in state space produced by model (\ref{I03}) might have some unknown effect on the
computation of  the SVDO coefficients. It is the first time that such coefficients are computed
from discontinuous data.
 % obsFN61_svd_delay.m

It is arguable that observabillity should not strongly depend on the dynamical regime. This view
which is something like ``structural observability'' is captured by the symbolic coefficients proposed
in \citep{let_agu/09pre}. On the other hand because the system visits different regions of the
state space during different dynamical regimes, and since  the
observability matrix might become singular at certain places, it can be expected that the
dynamical regime might have some influence on observability features.

The results concerning observability are summarized in Table~\ref{tb}.

\begin{table}[htb]
  \centering
  \caption{Observability coefficients for the different models investigated 
in this work.\label{tb}}
  \begin{tabular}{cccc}
%   ~\\[-0.3cm]
    \hline \hline
    & \multicolumn{3}{c}{Hodgkin-Huxley model} \\
    & $\delta_s$ & $\eta_{s^4}$ & $S_s$ \\[0.1cm]
    \hline
    ~\\[-0.4cm]
    $V$ & $1.1122 \cdot 10^{-7}$ & 0.12 & 0.109 \\
    $n$ & $1.0872 \cdot 10^{-6}$ & 0.19 & 0.093 \\
    $m$ & $5.1427 \cdot 10^{-9}$ & 0.19 & 0.054 \\
    $h$ & $4.8986 \cdot 10^{-6}$ & 0.19 & 0.167 \\[0.1cm]
    \hline
     ~\\[-0.4cm]
    & \multicolumn{3}{c}{FitzHugh-Nagumo model} \\
    & $\delta_s$ & $\eta_{s^2}$ & $S_s$ \\[0.1cm]
    \hline
     ~\\[-0.4cm]
    $x$ & Fig.~\ref{FNdeltas} & 1.00 & 0.301 \\
    $y$ & $0.09$ & 1.00 & 0.206 \\
    \hline
     ~\\[-0.4cm]
    & \multicolumn{3}{c}{Hindmarsh-Rose model} \\
    & $\delta_s$ & $\eta_{s^3}$ & $S_s$ \\[0.1cm]
    \hline
     ~\\[-0.4cm]
    $x$ & $3.16 \cdot 10^{-4}$ & 0.25 & 0.665 \\
    $y$ & $3.16 \cdot 10^{-4}$ & 0.56 & 0.449 \\
    $z$ & $3.16 \cdot 10^{-6}$ & 1.00 & 186.630 \\
    \hline
     ~\\[-0.4cm]
    & \multicolumn{3}{c}{Izhikevich's model} \\
    & $\delta_s$ & $\eta_{s^2}$ & $S_s$ \\[0.1cm]
    \hline
     ~\\[-0.4cm]
    $u$ & $0.1459$ & 1.00 & 0.750 \\
    $v$ & $0.3948$ & 1.00 & 8.376 \\[0.1cm]
    \hline \hline
  \end{tabular}
\end{table}

%=======================================================
\section{Synchronization in Networks of Neuron Models}
\label{snnm}

In this section, we provide numerical evidence that phase synchronization in networks of neuron models can be detected without estimating the phase, by using the svM-SSA with the single variable approach (see Sec.~\ref{mssa}). This is 
viable, and relevant in practice, due to the fact that the variable 
experimentally measured provides good observability of the system dynamics 
(Sec.~\ref{nr}).

This is illustrated by three examples, with increasingly complexity.
 The first  explored scenario is a chain of {\it phase coherent} FitzHugh-Nagumo neuron models. This allows one to compare the svM-SSA results with the one provided by the mean observed frequencies analysis (estimated through the phases of the analytical signal).
 The second scenario is the synchronization of coupled {\it bursting} neurons, which represents a challenging system with two very different time scales. For this, we investigate a chain of coupled Hindmarsh-Rose neuron models in the ``random'' bursting regime.
 The last example explores the feasibility of the svM-SSA itself in the synchronization analysis of coupled neurons in a {\it large} network. The svM-SSA provides detailed information of synchronization mechanism (e.g., PS clustering) in small networks, but what kind of information this technique could bring in large ones? This is investigated  in the context of a network of $1000$ Izhikevich's spiking neuron models. 
%

%-----------------------------------------------------------------------------------
\subsection{FitzHugh-Nagumo model}
\label{mssa.FH}

Consider  a chain of $J=5$ diffusively coupled FitzHugh-Nagumo neuron models:
\begin{eqnarray}
\label{FN61.coupled}
\left\{ 
\begin{array}{l}
\dot{x}_j = \displaystyle c(y+x-\frac{x^3}{3}+z) +C\sum_{i\in\Gamma_j}(x_i-x_j)  \\
\dot{y}_j= \displaystyle -(x-a+by)/c \\
\end{array} \right.
\end{eqnarray}

\noindent
with $j=1, ..., J$, where $C$ is the coupling strength, and $\Gamma_j$ is the set of values of $i$ that correspond to the oscillators coupled to $j$. Along this section, data was generated by integrating (\ref{FN61.coupled})  with integration step $h=0.01$ time units for a total time $t_{\rm sim}=450$ t.u., using a 4th-order Runge-Kutta algorithm. The first $t_{\rm trans}=50$ t.u.  were discarded, and the time series of $x$ and $y$ were sampled with sampling time $t_{\rm s}=0.7$ t.u.  (yielding approximately $15$ data points per period). The other parameters were  $(a,\,b)=(0.7, 0.8)$, with a detuning introduced by setting  $c_j=c_1+(j-1)\Delta c$, with $c_1=3$, $\Delta c=0.2$ and $I=-4$.

\begin{figure}[h]
\begin{center}
\includegraphics[width=0.7\textwidth]{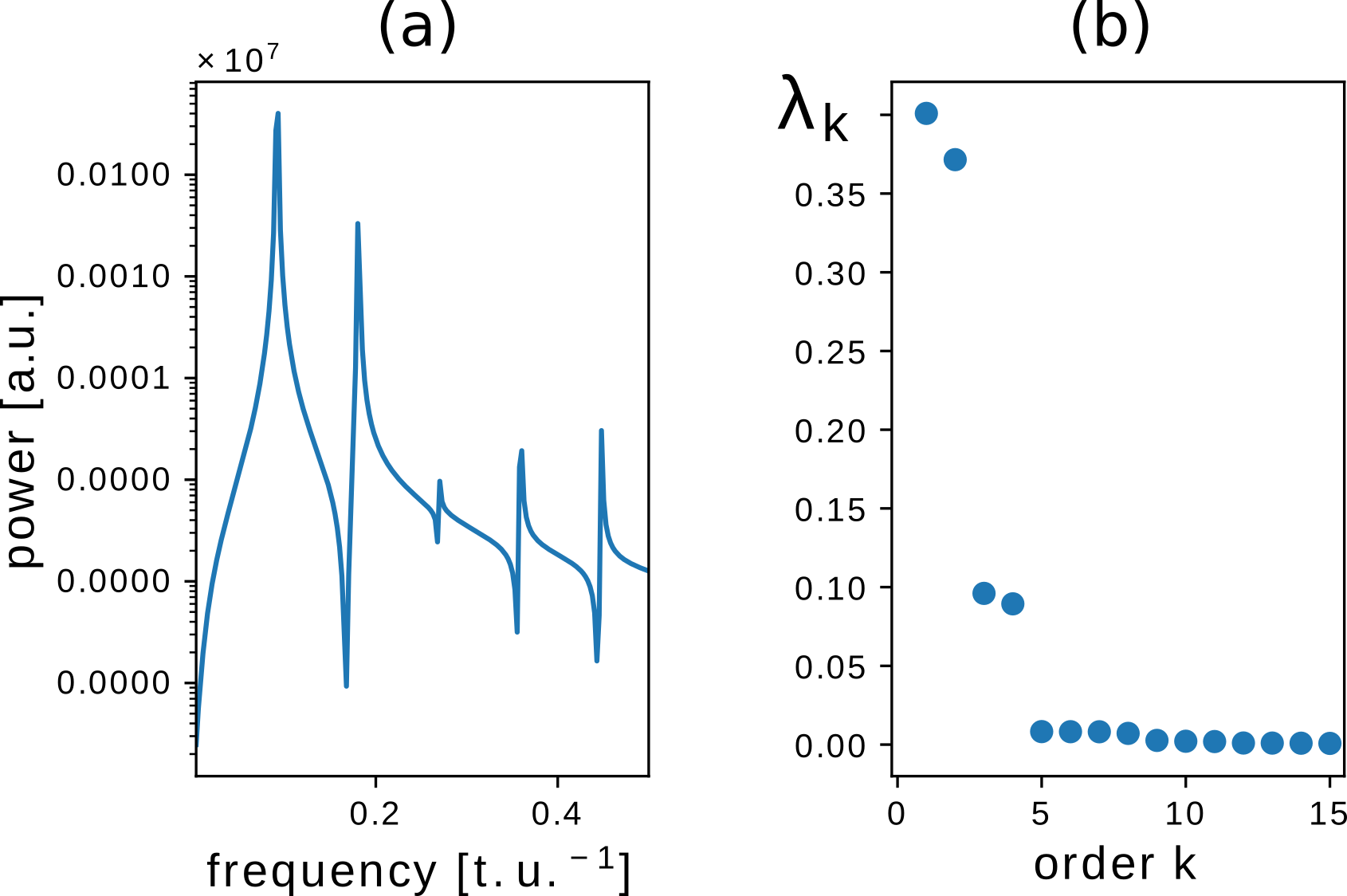}
\caption{\label{fig.FHN62.PSD}(a) Power spectrum density and (b) svM-SSA template analysis of a FitzHugh-Nagumo neuron. Two oscillatory modes are predominant, a stronger ($\lambda_{1,2}$) and a weaker one ($\lambda_{3,4}$).}
\end{center}
\end{figure}

Before starting the synchronization analysis, one needs to know the specific ``fingerprint'' of the oscillatory dynamics of one oscillator in the svM-SSA. Figure~\ref{fig.FHN62.PSD} shows the power spectrum of the $x$ time series  and the $20$  leading svM-SSA  singular values (the template for svM-SSA) for a single (uncoupled) neuron model ($j=1$). The signal present several harmonics, with the fundamental frequency at $\approx0.09$ Hz (a period of $T\approx11$ t.u.). Then, we set the svM-SSA's  window width  $m=31$ ($\approx2T/t_{\rm s}$), covering 
almost two oscillatory periods. The template in Fig.\,\ref{fig.FHN62.PSD}(b)
shows that a single oscillator will be identified by the svM-SSA as two leading pairs of singular values ($\lambda_{1,2}$ and $\lambda_{3,4}$), which are associated with the two strongest oscillatory modes of the signal. 

% data: FHN61-CHAIN-Nsys[5]-coup_vector[1, 0]-Delta_c[0.2]-C0-Cf-N[0, 0.2, 400]-s_time[0.1]
% notebook for plot: FHN61-mssa-template-and-PDS
%

% Experiment
We illustrate the use of the svM-SA by investigating the synchronization dynamics of the chain for an increasing coupling strength $C\in[0, 0.06]$. For the sake of clarity, a crosscheck of the results was done 
by considering  the mean frequency locking $\Omega_j=\Omega_i$ as a (weak) condition for the characterization of PS, and by the visual inspection of the spatiotemporal patterns $x_j(t)$ and $y_j(t)$ (as is usual in neuroscience). The $\Omega_j$ were computed through the linear least squares fit of the instantaneous phase $\phi_j(t)$ (estimated through the analytic signal based on the Hilbert transform).

\begin{figure}
\begin{center}
\includegraphics[width=0.7\textwidth]{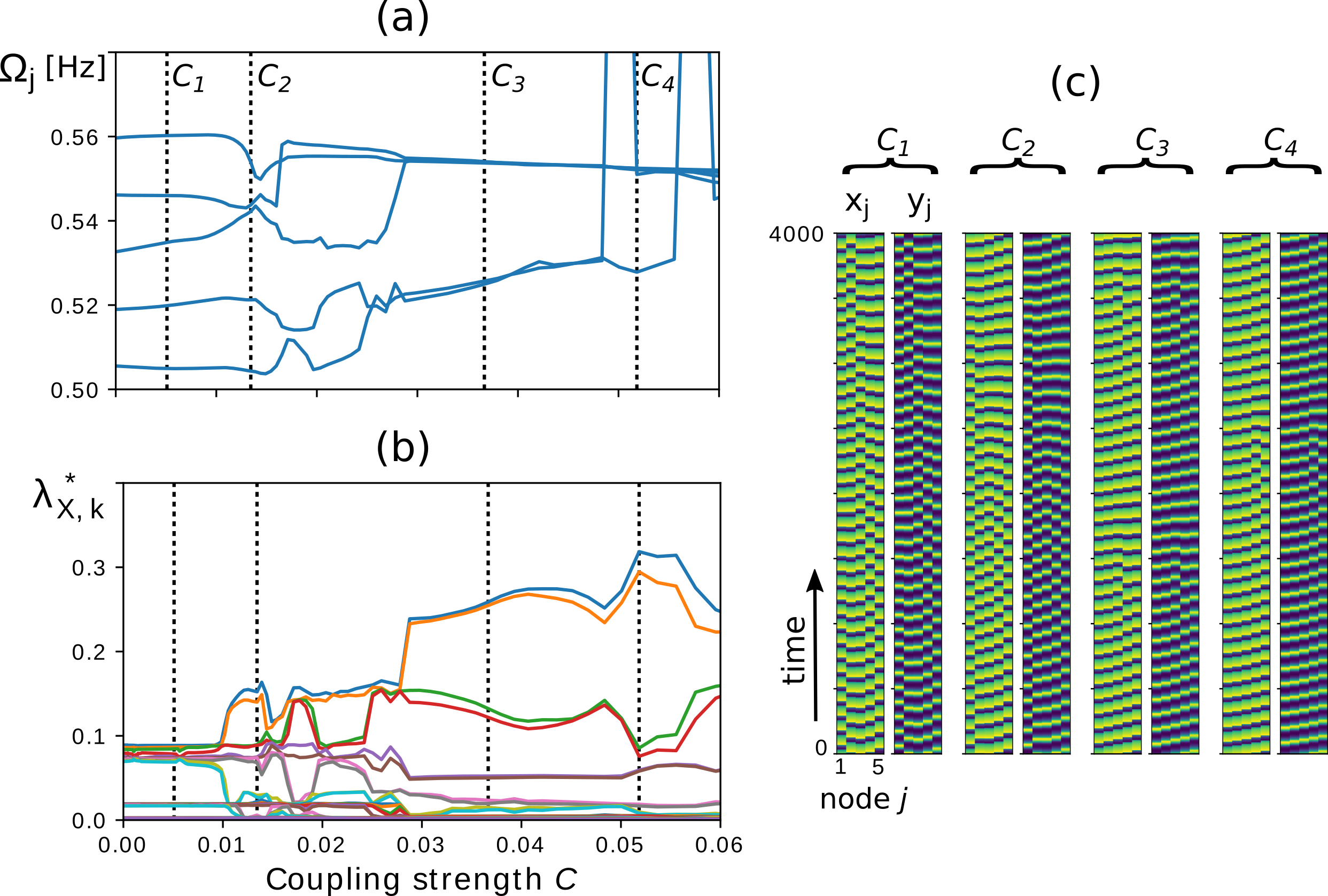}
\caption{\label{fig.FHN.mssa.chain}  Phase synchronization analysis of $J=5$   detuned FitzHugh-Nagumo neurons for an increasing coupling strength $C$. Four values of $C$ (vertical dashed lines) were selected to guide the following analysis. Both the (a) mean observed frequencies $\Omega_j$ (linear fit of the phases estimated from the analytical signal) and the (b) svM-SSA show the PS clustering formation. But the latter  (i)~provides an early sign of PS and (ii)~is not affected by phase slips as the former (as seen in the abrupt jumps of $\Omega_j$ at $C_4$). (c) A direct visual inspection of the $x$ and $y$ time series agrees with the results of these techniques, but its analysis is far less clear. }
\end{center}
\end{figure}

Figure~\ref{fig.FHN.mssa.chain} shows the results. The following features are worth noticing. 
First, at low values of the coupling strength (e.g., at $C=C_1$, first dashed vertical line) the five oscillators have different frequencies $\Omega$, suggesting no PS. Agreeing with that, the svM-SSA is equivalent to the ``concatenation'' of $5$ individual $\lambda$ templates.
 Second, the onset of phase synchronization, identified by the mean frequency 
analysis as $\Omega_3=\Omega_4$, occurs at $C_2$ (second dashed vertical line). 
The svM-SSA captured the onset of PS much earlier, identified by the increasing 
value of a $\lambda^*$ pair with a simultaneous drop of other one to the noise 
floor.
 Third, several ``jumps'' are presented in the mean frequency plot, associated with episodes of poor or inappropriate frequency estimates (e.g., due to phase slips), one of them at $C=C_4$. We obtained this result regardless the phase estimate definition used (i.e, $\arctan(y/x)$ or through a Poincar\'e section, not shown). Since the svM-SSA does not require the computation of phases, no ``jumps'' are present. This robustness represents another advantage of the method. 
 Fourth, the increasing synchrony suggested by spatiotemporal patterns, at four illustrative values of $C$, agrees well with  the svM-SSA.
 Other features captured by the svM-SSA are present, but are out of the scope of the present work. The main message of the aforementioned results is that svM-SSA provides detailed information about the synchronization dynamics.

% data: FHN61-CHAIN-Nsys[5]-coup_vector[1, 0]-Delta_c[0.2]-C0-Cf-N[0, 0.2, 400]-s_time[0.1]
% Notebooks for plots: FHN61-mssa-for-PAPER, FHZ-mean-observed-frequency, Spatiotemporal-patterns

%----------------------------------------------------------------------------
\subsection{Hindmarsh-Rose model}
\label{mssa.HR}

In this section we consider a network of identical neuron models but with different initial conditions in a chain topology of $J=5$ oscillators. The dynamical regime of such models ($I=3.25$) lead to chaotic bursts (see Sec.~\ref{mod_HR}).

\begin{figure}[h]
\begin{center}
\includegraphics[width=0.7\textwidth]{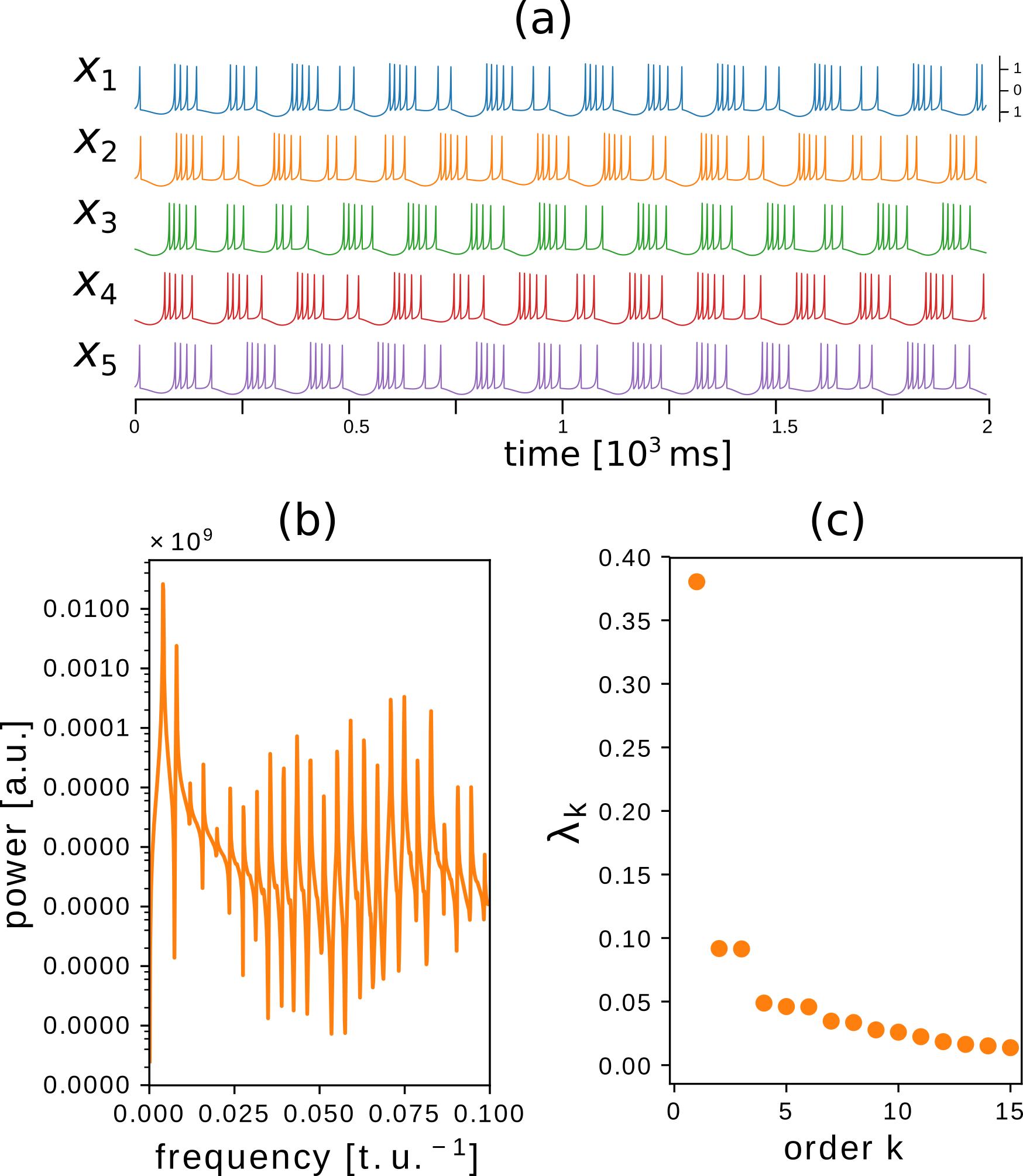}
\caption{\label{fig.HR84.PSD}(a) $x$ time series of the ({\it uncoupled}) $J=5$ simulated Hindmarsh-Rose neurons show the generated ``random'' bursting structures. For a given oscillator, the (b) PSD shows two dominant low frequencies (from the slow oscillatory mode), and the (c) template for the svM-SSA displays a ``drift'' signature (leading isolated $\lambda_1$, see text) and a slowly decreasing tail of singular value pairs corresponding to the several oscillatory modes present in the signal.}
\end{center}
\end{figure}

Consider a set of HR neuron models coupled according to \citep{han_sha/16} 
\begin{eqnarray}
\label{eq.HR84.coupled}
\left\{ 
\begin{array}{l}
\dot{x_j}  =  \displaystyle y_j -ax_j^3+b x_j^2+I -z_j  +g\sum_{i\in\Gamma_j}(x_i) \\
\dot{y_j}  =  \displaystyle c - d x_j^2 -y_j  \\
\dot{z_j}  =  \displaystyle r[s(x_j-x_1)-z_j] ,  
\end{array} \right.
\end{eqnarray}
with $j=1, ..., J$, and $\Gamma_j$ being the set of values of $i$ that correspond to the oscillators coupled to $j$. Data was generated by integrating (\ref{eq.HR84.coupled}) with $(h, t_{\rm s},t_{\rm sim}, t_{\rm trans})=(0.1,0.1,8500,500)$ -- for the svM-SSA, data was decimated by $10$, yielding an effective $t_{\rm s}=1$ t.u. We set $(a,b,c,d)=(1,3,1,5)$ and $I=3.25$, in order to generate a ``random'' burst structure. Figure~\ref{fig.HR84.PSD}(a) shows that each neuron has a different bursting sequence for the coupling strength $g=0$ (i.e., uncoupled). 

The ``period'' of the fast time scale corresponds to the inter-spike interval in a burst, which is $\approx15$ data points (in the decimated time series). Hence, we set $m=30$ following \citep{por_agu/16chaos}. In doing so, a  single high $\lambda$ [Fig.~\ref{fig.HR84.PSD}(c)] is seen in the svM-SSA template. It is followed by a slowly decreasing tail of singular pairs, which correspond to the several time scales present in the signal [the PSD, \ref{fig.HR84.PSD}(b), shows two of them]. Finally, we choose $S=25$ eigenvectors of the structured-varimax rotation, which correspond to the  $5$ leading singular values of each  single-neuron as its phase dynamics fingerprint (based on the template).

% data: DATA_HR Neuron model_Nsys5_CoupStr[0]Delta_a0.016[Stime,TotalTime][0.1, 8000]
% notebook for plot: HR84-mssa-template-and-PDS

\begin{figure}[h]
\begin{center}
\includegraphics[width=0.7\textwidth]{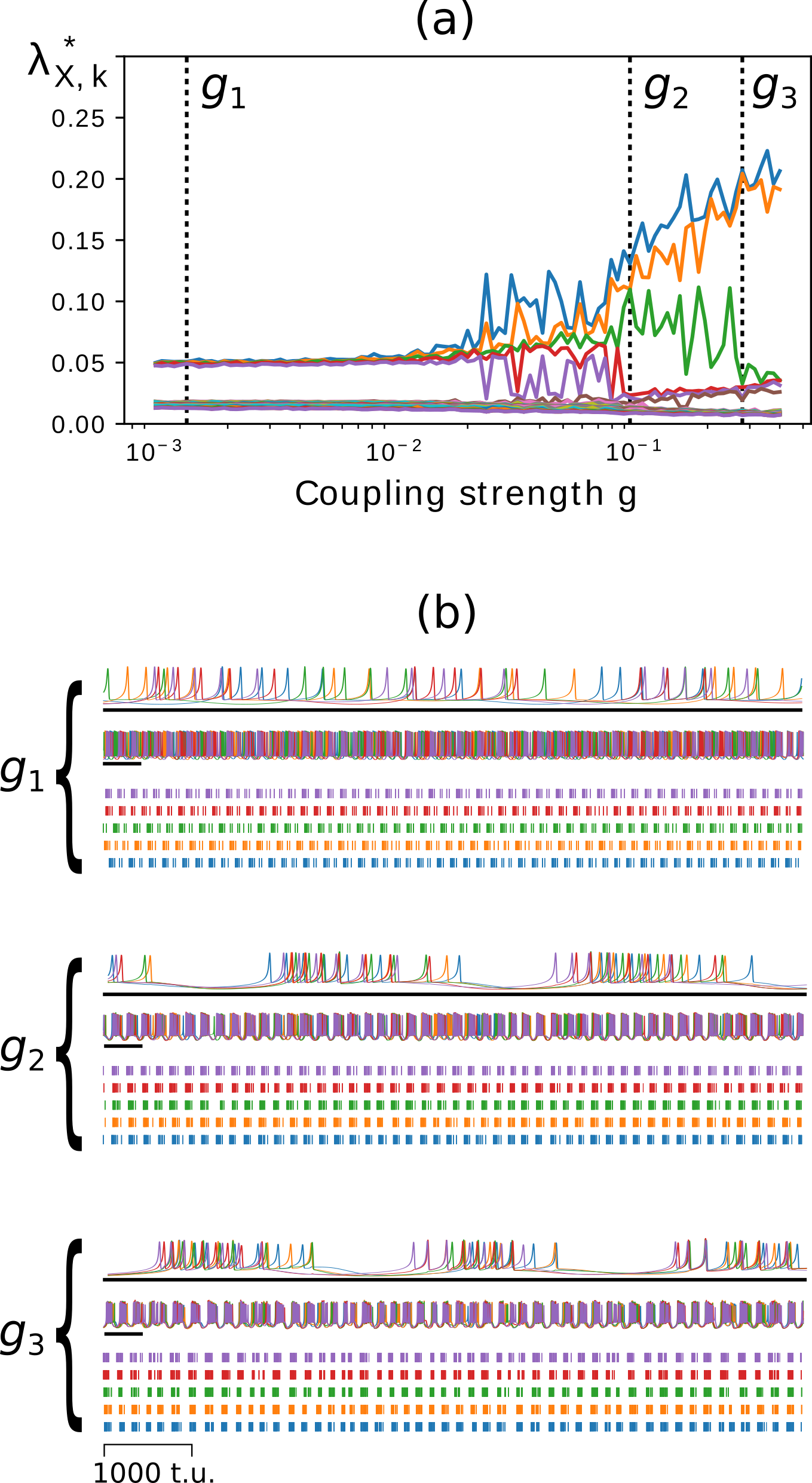}
\caption{\label{fig.HR84.mssa}Synchronization analysis of $J=5$ Hindmarsh-Rose neurons coupled in a chain, for an increasing synaptic coupling strength $g$. (a) the svM-SSA shows the PS dynamics. For $g\approx 0$, each bursting neuron is identified by its single leading $\lambda_k$ (as suggested by the template analysis, see text). Increasing $g$, the spectrum suggests an increasing synchronization due the formation of PS clusters. (b) the $x$ time series at three selected values of $g$ (vertical dashed lines), along with a ``zoom'' view of the initial segment, show an increasing tendency to synchronization, in agreement with the svM-SSA results. Intermittent synchronization can be seen in the raster plots.}
\end{center}
\end{figure}

Figure~\ref{fig.HR84.mssa} shows the results for $100$ (logarithmically spaced) steps of an increasing synaptic coupling strength $g\in [0,0.4]$. For $g\approx 0$, each neuron is represented by the corresponding leading $\lambda_1$. Increasing $g$,  a somewhat intermittent PS appears. Three values of $g$ (vertical dashed lines), that reveal increasing PS, were selected. The respective time series, Fig.~\ref{fig.HR84.mssa}(b), confirm the increasing synchronization 
of the bursts. Note that the spikes themselves are not synchronized (as seen in the selected initial segments and in the respective raster plots). The two high singular values for $g=g_3$ indicate PS behavior of two dominant clusters.
From the raster plots it is seen that such clusters are not fixed, in the sense that some neuron models synchronize
intermittently with each of them. The detailed analysis of this intermittent PS behavior is left for future research.

% data: DATA_HR Neuron model_Nsys5_CoupStr[0]Delta_a0.016[Stime,TotalTime][0.1, 8000]
% notebook for plot: HR84-mssa-template-and-PDS
%

%----------------------------------------------------------------------------
\subsection{Izhikevich's spiking neuron model}
\label{mssa.IZ}

As seen in the previous section, the svM-SSA is a powerful tool to provide information about phase synchronization dynamics, at least for {\it few}\, oscillators. But computational investigations on neuronal dynamics also include networks with a {\it massive}\, number of neurons such as  $J\thicksim 10^3$ up to $10^6$ and above.  For example, a network model of the mammalian thalamocortical system exhibited polychronous activity only with $J>10^4$ neurons \citep{izh_ede/08}.

Hence, in this section we explore what kind of information, if any, the svM-SSA can provide about synchronization dynamics in a large population of neuron models. To this end, a numerical experiment using a network with $J=1000$ Izhikevich neuron models is reported. Such a model has both biological plausibility and computational efficiency, been able to emulate nearly $20$ neuro-computational properties of biological spiking neurons and hence is adequate for this type of simulation studies \citep{izh/04}. The same rationale (parameters, integration method etc) of the original work \citep{izh/03} was followed, unless otherwise stated.
%

\begin{figure}
\begin{center}
\includegraphics[width=0.7\linewidth]{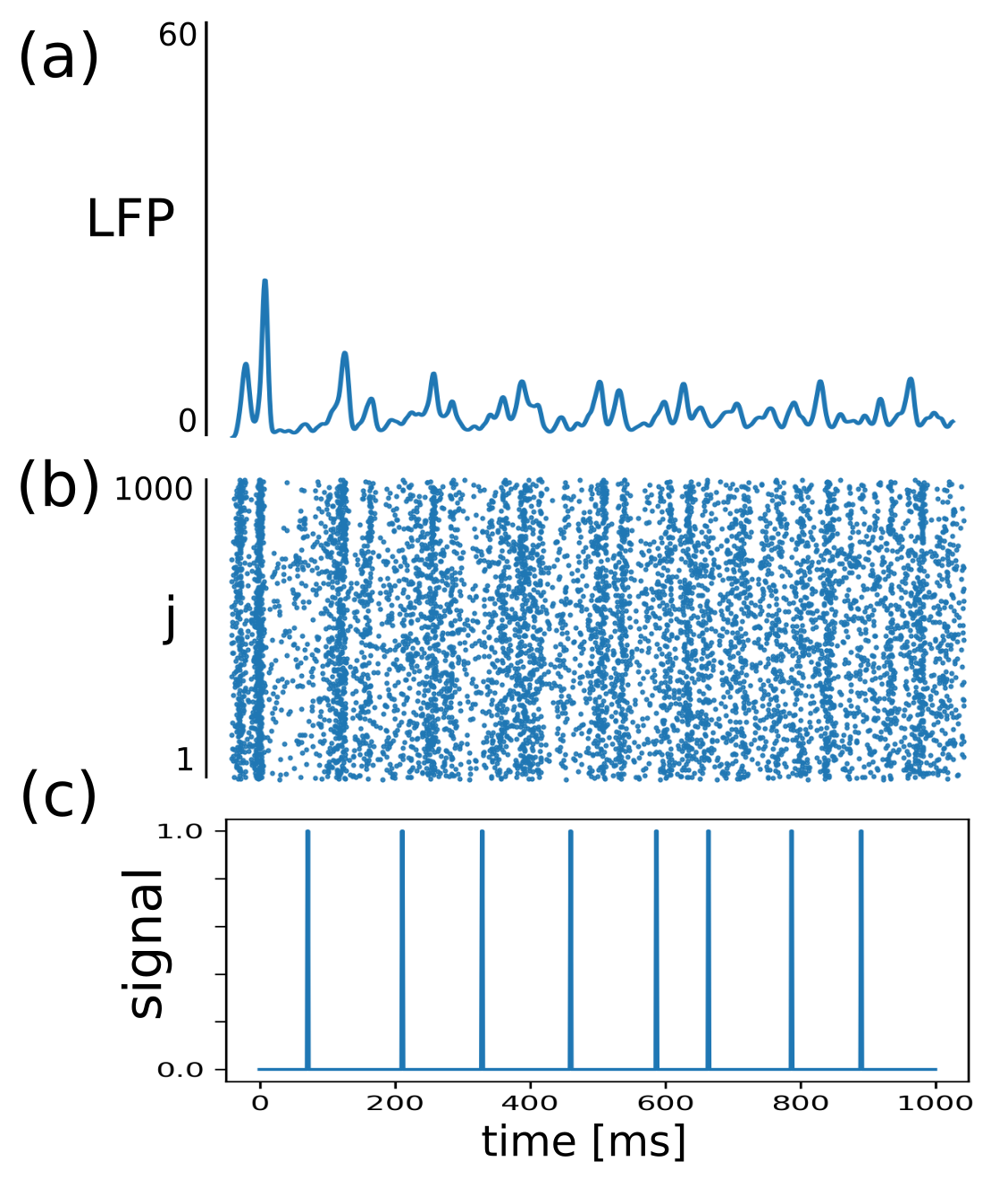}
\caption{\label{fig.iz03lfp}Simulation results for $g=1$. 
(b) Spike raster and (a) its respective local field potential, showing 
signatures of spiking synchronization. (c) Typical spike raster signal, 
obtained from neuron $j=1$.}
\end{center}
\end{figure}

% The experiment
We investigate the synchronizability of the network in function of the coupling strength. This is done by manipulating the synaptic current ($I_{\rm syn}$) with an scaling factor (coupling gain) $g\in[0, 1]$. This is implemented by splitting the input current term in (\ref{I03}) as $I^j=I_{\rm in}^j+gI_{\rm syn}^j$ (being $I_{\rm in}$ the injected dc-current, and $j=1, ..., J$ the neuron index). The {\it same} randomly generated synaptic weight matrix  was used in all simulations, as well the number of excitatory ($N_{\rm e}=800$) and inhibitory ($N_{\rm i}=200$) neurons. For a given value of $g$,  the {\it spike raster}\, time series were generated with $(h, t_{\rm s},t_{\rm sim}, t_{\rm trans})=(1,1,1000,0)$ (ms time units). Figure~\ref{fig.iz03lfp}(a-c) shows a representative example for $g=1$, along with the respective local field potential. One hundred values of $g$ were used, equally spaced in the aforementioned range.

Note that the relevant dynamical information is coded in the inter-spike interval (ISI), and not in their magnitude. In view of this, the raster time series from $v(t)$ is used in order to decrease the computational load from the augmented trajectory matrix. 
Following \citep{por_agu/16chaos} for the svM-SSA of the Hindmarsh-Rose neuron model (see Sec.~\ref{mssa.HR}), the window width was chosen as $m=4$, that results in a window length of $4\times t_{\rm s}=4$ ms, which is four times larger than the fast time scale ($1$ ms of the spike duration). For the structured varimax rotation only the $S=2J= 2000$ leading singular vectors were used which requires the rotation of a matrix of size $2000\times 2000$.\footnote{This is done for two reasons. First, in order to minimize computational effort, since the structured varimax rotation algorithm \citep{por_agu/16pre}  is based on a singular value decomposition (SVD), which is known to have a time complexity of order $O[{\rm min}(pq^2,p^2q)]$ for a generic matrix of size $p\times q$ (i.e., $O(2000^3)$ in the present case). Second, to simplify the analysis of the svM-SSA spectrum by following just one oscillatory mode per system, corresponding to two singular values $\lambda_k$.}

% IZ03-g0-gf-Ng-[0, 1, 100]NeNi-[800, 200].npy

\begin{figure}
\begin{center}
\includegraphics[width=0.7\linewidth]{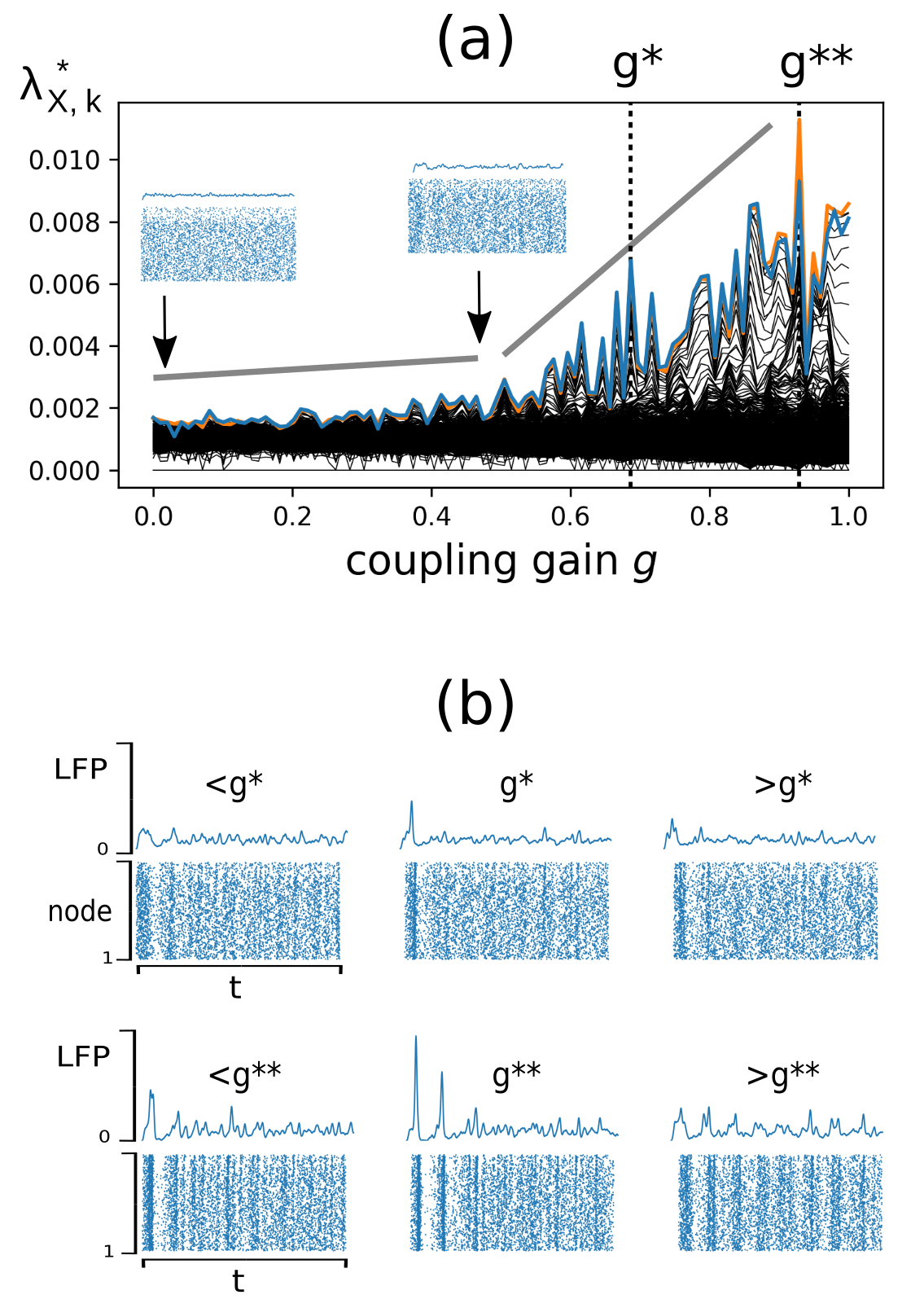}
\caption{\label{fig.IZ-mssa-G}(a) (a) Modified singular values $\lambda_{Y,k}^*$ for $N=1000$ pulse-coupled Izhikevich neuron models for an increasing inter-neuron coupling gain $g$. The two thick (blue and orange) lines correspond to the highest singular values, $\lambda^*_{1,2}$.
 The increasing in synchronicity, for larger values of $g$, have a higher slop (thick gray lines) for $g>0.5$ The insets illustrate typical spike raster plots (SRP) and their respective local field potential (LFP) found for $g<0.5$. No clear signature of synchronicity is seen, which agrees with the low values of $\lambda_k$ for this range of $g$.
  Two high peaks in the svM-SSA spectrum ($g^*$ and $g^{**}$), suggesting PS, were selected for closer inspection. Their respective LFP and SRP, along with the ones for the surrounding values of $g$, are shown in (b). They show a clear sign of synchronism, which is stronger near $g^{**}$ than near $g^*$, in agreement with the svM-SSA spectrum.}
\end{center}
\end{figure}

The svM-SSA spectrum for the increasing coupling gain $g$ is shown in Fig.~\ref{fig.IZ-mssa-G}(a). The 
behavior and quantity of the $\lambda_k^*$  render the plot complicated and no {\it specific} phase synchronized 
cluster information is clear seen. On the other hand, a {\it general} picture of the synchronization dynamics is 
provided if one focuses on  $\lambda_{1,2}^*$ (the leading pair), corresponding to the stronger, global, oscillatory 
mode present in the data. The following features are worth mentioning.
First, there is a clear overall growing trend of $\lambda_{1,2}^*$ values for increasing gain $g$ (marked by the gray thick lines).
Second, the spectrum suggests that no global PS emerges until $g\approx 0.5$, where the slope of the aforementioned trend starts to increase. 
The insets in Fig.~\ref{fig.IZ-mssa-G}(a) show the spike raster and respective 
local field potential (LFP), for two representative values of $g<0.4$, 
with no clear visible sign of synchronized spiking. 
Third, higher values of $g$ do not necessarily imply a ``higher level'' of PS, as shown by the presence of high peaks (higher level of PS) surrounded by deep valleys (lower level of PS).  As a cross-check, the raster spikes and respective LFP at and around the two selected peaks at $\{g^*, g^{**}\}=\{0.6868, 0.9292\}$ are shown in Figs.~\ref{fig.IZ-mssa-G}(b,\,c),
respectively. Both ranges show signatures of synchronization, which is more visible near the second (and higher) peak at $g=g^{**}$ [Figs.~\ref{fig.IZ-mssa-G}(c)]. 
Hence in the context of a large network, the svM-SSA is still able to provide the general picture of PS.

% IZ03-g0-gf-Ng-[0, 1, 100]NeNi-[800, 200].npy
%

%=======================================================
\section{Conclusions}
\label{conc}

In most studies involving neuron models, it is common to use the first state 
variable -- the membrane potential -- for monitoring or controlling purposes. 
This choice of variable results from the fact that, in experimental
neuron network, only this variable can be actually measured. However, it is important, from a theoretical point
of view, to know if such a choice is the most adequate in terms of the dynamical behavior.

One of the objectives of this paper has been to investigate observability properties of 
neuron models. This has been done using three different quantifiers for observability:
coefficients determined numerically from the model equations \citep{let_eal/05pre},
from data \citep{agu_let/11} or symbolic coefficients analytically obtained from  
the model equations \citep{let_agu/09pre}. This procedure turned out to reveal the
limitations of some techniques, for instance, due to the complexity of the equations
and the physical interpretation of the variables, investigating the observability of the Hodgkin-Huxley
model is viable only using the data-estimated SVDO coefficients 
or the symbolic observability coefficients. Also, the performance of such a method using
discontinuous data as for Izhikevich's spiking neuron model is uncertain. 
This mostly results from the fact that in this latter model, the 
switching mechanism is not fully described by the equations and there is at 
least one missing variable in the model for having a complete description of
the underlying mechanisms. Observability is therefore investigated from a 
truncated model and it remains an open question how to assess observability
in such discontinuous systems.

In summary the variables that convey greater observability were: the membrane 
potential in the Hodgkin-Huxley and Izhikevich's models (especially in the 
chattering regime), whereas for FitzHugh-Nagumo the 
observability provided by the potential and recovery variables is comparable.
In this respect, the Hindmarsh-Rose model has some peculiarities in what 
concerns observability. The membrane potential and fast recovery variable
reveal the fast time scales such as the spikes in the chattering regime, 
whereas the $z$ variable (slow recovery) is the only one to clearly reveal the 
chaotic nature of the dynamics when it occurs. 

We also investigated some of the aforementioned models in the context of synchronization.
In particular, networks formed of five phase coherent FitzHugh-Nagumo neurons,
five bursting Hodgkin-Huxley neurons,  and one with $1000$
Izhikevich neurons were analyzed. A technique known as structured-varimax 
multivariate singular spectrum analysis, from a variable that provides good
observability of the dynamics, was used to successfully detect phase synchronization in 
the networks.
Two interesting features of this technique is that it does not require computing the phase
and it is able to detect synchronization in situations where other methods give an unclear
indication.

%=======================================================
\section*{Acknowledgements}

The authors gratefully acknowledge financial support from Conselho
Nacional de Desenvolvimento Cient\'ifico e Tecnol\'ogico (CNPq), Brazil.

\bibliographystyle{apalike}

\end{document}